\pdfoutput=1

\documentclass[11pt,nofootinbib]{article}
\usepackage{amsmath}
\usepackage{amssymb}
\usepackage{graphicx}
\usepackage[colorlinks=true,linkcolor=blue,citecolor=blue]{hyperref}

\numberwithin{equation}{section}

\newcommand{\bea}{\begin{eqnarray}}   
\newcommand{\eea}{\end{eqnarray}}
\newcommand{\bear}{\begin{array}}  
\newcommand {\eear}{\end{array}}
\newcommand{\bef}{\begin{figure}}  
\newcommand {\eef}{\end{figure}}
\newcommand{\bec}{\begin{center}}  
\newcommand {\eec}{\end{center}}

\def\EQ#1{Eq.~(\ref{#1})}

\def\GEV#1{10^{#1}{\rm\,GeV}}

\newcommand{\hyphen}{{\rm \,\mathchar`- \,} }

\begin{document}

\begin{titlepage}

\setcounter{page}{1} \baselineskip=15.5pt \thispagestyle{empty}

\begin{flushright}
ICRR-Report-633-2012-22\\
IPMU12-0190\\
TU-922
\end{flushright}
\vfil

\bigskip\
\begin{center}
{\LARGE \textbf{Non-Gaussianity from Axionic Curvaton}}
\vskip 15pt
\end{center}

\vspace{0.5cm}
\begin{center}
{\Large 
Masahiro Kawasaki,$^{a,b}$\footnote{kawasaki@icrr.u-tokyo.ac.jp}
Takeshi Kobayashi,$^{c,d}$\footnote{takeshi@cita.utoronto.ca}
and Fuminobu Takahashi$^{b,e}$\footnote{fumi@tuhep.phys.tohoku.ac.jp}
}
\end{center}

\vspace{0.3cm}

\begin{center}
\textit{$^{a}$ Institute for Cosmic Ray Research, The University of
 Tokyo, \\ 5-1-5 Kashiwanoha, Kashiwa, Chiba 277-8582, Japan}\\

\vskip 4pt
\textit{$^{b}$ Institute for the Physics and Mathematics of the
 Universe, The University of Tokyo, \\ 5-1-5 Kashiwanoha,
 Kashiwa, Chiba 277-8582, Japan}\\

\textit{$^{c}$ Canadian Institute for Theoretical Astrophysics,
 University of Toronto, \\ 60 St. George Street, Toronto, Ontario M5S
 3H8, Canada}\\ 

\vskip 4pt
\textit{$^{d}$ Perimeter Institute for Theoretical Physics, \\ 
 31 Caroline Street North, Waterloo, Ontario N2L 2Y5, Canada}

\vskip 4pt 
\textit{$^{e}$ Department of Physics, Tohoku University, Sendai 980-8578, Japan}
\end{center} \vfil

\vspace{0.8cm}

\noindent
We study non-Gaussianity of density perturbations generated by an axionic curvaton,
focusing on the case that the curvaton sits near the hilltop of the potential during inflation.
Such hilltop curvatons can generate a red-tilted density perturbation spectrum without
invoking large-field inflation. We show that, even when the curvaton dominates the Universe, 
the non-Gaussianity parameter $f_{\rm NL}$ is positive and mildly increases towards the hilltop of the curvaton potential, and that
$f_{\rm NL} = {\cal O}(10)$ is a general and robust prediction of such hilltop axionic curvatons.
In particular, we find that the non-Gaussianity parameter is bounded as $f_{\rm NL} \lesssim 30 \hyphen 40$
for a  range of the scalar spectral index, $n_s = 0.94 \hyphen 0.99$, and that
$f_{\rm NL} = 20 \hyphen 40$ is realized for the curvaton mass $m_\sigma = 10 \hyphen 10^6$\,GeV and the decay constant $f = 10^{12} \hyphen 10^{17}$\,GeV.
One of the plausible candidates for the axionic curvaton is an imaginary component of a modulus field 
with mass of order $10 \hyphen 100$\,TeV and decay constant of  $\GEV{16
 \hyphen 17}$. 
We also discuss extreme cases where the curvaton drives a second
inflation and find that $f_{\rm NL}$ is typically smaller compared to
non-inflating cases. 
\vfil

\end{titlepage}

\newpage
\tableofcontents

\section{Introduction}
\label{sec:intro}

Several theoretical difficulties of the standard big bang cosmology such
as the horizon and flatness problems
can be elegantly solved by inflation~\cite{Guth:1980zm}. In fact, 
the existence of the inflationary era in the early
Universe is strongly supported by the observations~\cite{Komatsu:2010fb}; 
the density perturbations extending beyond the horizon at the last
scattering surface can be interpreted as the evidence for the
accelerated expansion in the past.

The study of density perturbations such as isocurvature perturbations, non-Gaussianity,
tensor-mode, and their effects on the cosmic microwave background (CMB)
power spectrum is a powerful diagnostic of the mechanism that laid down the primordial density
fluctuations, but it is not enough at present to pin down the model.  This is partly because of our ignorance of thermal
history of the Universe beyond the standard big bang cosmology,
especially concerning  how the Universe was reheated. 

Whereas one of the plausible explanations for the density perturbations is the quantum fluctuations of the 
inflaton from the minimalistic point of view, it may be that there are many other light scalars in nature, 
one of which is responsible for the observed density perturbation via the curvaton~\cite{Linde:1996gt,Enqvist:2001zp,Lyth:2001nq,Moroi:2001ct} 
(or its variant, e.g. modulated reheating~\cite{Dvali:2003em,Kofman:2003nx}) mechanism.
In fact, there are many moduli fields that necessarily appear at low energies through 
compactifications in string theory. Most of them must be stabilized in order to have a sensible low-energy theory,
but some of them may remain relatively light, and therefore are a candidate for the curvaton. 
Interestingly, there is  an argument that string theory contains 
a plenitude of axions, the so called ``string axiverse."\cite{Arvanitaki:2009fg} 
We shall see later that the axion is indeed a plausible candidate for the successful curvaton.\footnote{
Throughout this paper the axions refer to imaginary components of moduli fields, which have a sinusoidal 
potential. 
}

One of the distinguishing features of the curvaton mechanism is that it can generate the density perturbation with 
large non-Gaussianity. If any primordial non-Gaussianity is found by the Planck satellite, it would exclude
a simple class of inflation models as the origin of the entire density perturbation, and therefore, it has a
tremendous impact on our understanding of the early Universe. 

Recently, the present authors studied  non-Gaussianity generated by the curvaton mechanism in great detail,
and developed a formalism to calculate the density perturbation for a generic curvaton potential~\cite{Kawasaki:2011pd}.
We pointed out that the curvaton  should be located at a potential with negative curvature during inflation,
and in particular it must be close to the local maximum (``hilltop") of the potential,
in order to generate a red-tilted density perturbation spectrum which is strongly favored by the recent observations~\cite{Komatsu:2010fb}. 
Interestingly, we found that, even if
the curvaton dominates the Universe,  the 
non-Gaussianity parameter $f_{\rm NL}$ is positive and gets enhanced logarithmically in the hilltop limit, and therefore
$f_{\rm NL}$ of ${\cal O}(10)$ is a robust prediction of the hilltop curvaton. 
 Applying our formalism to the axionic (or pseudo-Nambu-Goldstone) curvaton with a sinusoidal potential, we found that $f_{\rm NL}$ can be as large as 
 about $30$, which is realized for the curvaton mass of order $10$ TeV and the decay constant of order the GUT scale. 
In this analysis we fixed the scalar spectral index $n_s = 0.96$ for simplicity. 
The mild increase of the non-Gaussianity in the hilltop limit is originated from the fact that
the density perturbation generated by the curvaton is enhanced. This enhancement is due to
non-uniform onset of curvaton oscillations~\cite{Kawasaki:2008mc,Kawasaki:2011pd}.
This result should be contrasted to
a simple curvaton model with a quadratic potential, which predicts a negative $f_{\rm NL}$ of order unity in the case
that it dominates the Universe.

In this paper, we extend our previous work on the non-Gaussianity
generated by the axionic curvaton with the hilltop initial condition.
We will discuss its dependence on the scalar spectral index, and also scan the curvaton parameters, 
namely, the mass and the decay constant. Interestingly, we find 
that $f_{\rm NL}$ is bounded as $f_{\rm NL} \lesssim 30 $ for $n_s = 0.94 {\rm \,\mathchar`- \,} 0.99$, and 
 the maximal non-Gaussianity is realized for the curvaton mass $10$ TeV
 and the decay constant of order the GUT scale. (If reheating happens
 prior to the curvaton oscillation, then the bound becomes
 $f_{\mathrm{NL}} \lesssim 40$.)
Furthermore,  $f_{\rm NL} = 20 \hyphen 40$ is realized for a wide range of parameters,
the curvaton mass $m_\sigma = 10 \hyphen 10^6$\,GeV and the decay constant $f = 10^{12} \hyphen 10^{17}$\,GeV.
One of the plausible candidates for such an axionic curvaton is an imaginary component
of the moduli (i.e., axions)  with mass of order $10 \hyphen 100$\,TeV and decay constant of  $\GEV{16 \hyphen 17}$.
The moduli fields are stabilized by the
non-perturbative effect and the supersymmetry (SUSY) breaking, and it is plausible that the moduli mass is closely related to
the SUSY breaking scale in the visible sector.  Intriguingly, such several tens TeV SUSY breaking scale
is consistent with the recently discovered Higgs boson mass of $125 \hyphen 126$\,GeV~\cite{:2012gk,:2012gu}.

The rest of the paper is organized as follows. After briefly reviewing 
density perturbations from general curvatons in
Section~\ref{sec:review}, then in Section~\ref{sec:pNGcurvaton} we
discuss axionic curvatons in detail. We then give discussions and
conclusions in Section~\ref{sec:Disc} and~\ref{sec:Conc},
respectively. 

The appendix discusses an extreme case where the curvaton
drives a second inflationary period. After analytically computing
density perturbations from inflating curvatons in general, we then apply
the discussions to axionic curvatons. We find that the non-Gaussianity
turns out to be rather small when the axionic curvaton drives a second inflation.

\section{Review of Curvatons with a General Potential}
\label{sec:review}

In the curvaton mechanism, the light curvaton field acquires
super-horizon field fluctuations during inflation. 
The density perturbations are produced in the post-inflationary era, 
as the curvaton oscillates and its energy
density relatively grows compared to other radiation components.
In this section we give a brief review of density perturbations
generated by a curvaton~$\sigma$ with a generic effective
potential~$V(\sigma)$. 
We refer the reader to~\cite{Kawasaki:2011pd} for detailed derivation of
the following results. 

\subsection{Density Perturbations from Curvatons}
The density perturbations generated by curvatons depend on the curvaton dynamics
during and after inflation. In the simple curvaton model with a quadratic potential,
the curvaton dynamics is determined by the curvaton mass and the initial deviation
from the origin. If the  mass is much smaller than the Hubble parameter during inflation, 
the curvaton hardly evolves until it starts to oscillate, and  the resultant
density perturbation is given in a rather simple form.  However this is no longer the case for a general
curvaton potential. 
In particular, the curvature of the potential should be negative and
non-negligible in order to account for the observationally favoured
red-tilted perturbation spectrum, then the curvaton
significantly evolves after inflation, affecting the density perturbation. 

If the curvaton potential~$V(\sigma)$ has no explicit dependence on time,
then the curvaton dynamics prior to the oscillation can be
tracked by the attractor solution
\begin{equation}
 \hat{c} H \dot{\sigma} = -V', \qquad
\mathrm{with} \quad
 \hat{c} = 
 \left\{
   \begin{array}{cl}
     3 & \mbox{(during inflation with $H \simeq $ const.)} \\
     9/2 & \mbox{(matter domination) } \\
     5 & \mbox{(radiation domination) }
   \end{array}
\right.
 \label{CDapp}
\end{equation}
which is a good approximation while $|V'' / \hat{c} H^2 | \ll 1$.
Here, a prime denotes a derivative with respect to~$\sigma$, an
overdot a time derivative, and $H = \dot{a} / a$. Setting the minimum of
the potential about 
which the curvaton oscillates to $\sigma =0$, the onset of the
oscillation can be defined as when the time scale of the curvaton
rolling becomes comparable to the Hubble time, i.e. 
\begin{equation}
 \left| \frac{\dot{\sigma}}{H\sigma} \right| = 1. 
\label{conditionX}
\end{equation}
Then the Hubble parameter at the time is obtained as
\begin{equation}
 H_{\mathrm{osc}}^2 = \left|
\frac{V'(\sigma_{\mathrm{osc}})}{c
  \sigma_{\mathrm{osc}}}
\right|,
 \label{Hosc}
\end{equation}
where the subscript ``osc'' denotes values at the onset of the
curvaton oscillation, and $c$ is a constant depending on 
whether reheating (= inflaton decay, at~$t_{\mathrm{reh}}$) is
earlier/later than the onset of the curvaton oscillation (corresponding
to $\hat{c}$ in the attractor~(\ref{CDapp}) right before the oscillation):
\begin{equation}
 c = 
 \left\{
   \begin{array}{cl}
        9/2 &  (t_{\mathrm{reh}} > t_{\mathrm{osc}}) \\
     5 & (t_{\mathrm{reh}} < t_{\mathrm{osc}}).
   \end{array}
\right.
 \label{ciroiro}
\end{equation}
The absolute value sign in (\ref{Hosc}) can be removed by supposing the
curvaton potential to be monotonically increasing (decreasing) for
$\sigma > (<) 0$, so that the curvaton can roll down to the origin.

Let us here summarize simplifying assumptions concerning the evolution
of the energy densities of the curvaton and the inflaton.
We assume the curvaton potential to be well approximated by
a quadratic one around its minimum so that the curvaton oscillations are
sinusoidal.\footnote{Cases with non-sinusoidal oscillations are
discussed in Appendix~B of~\cite{Kawasaki:2011pd}. We note that 
the oscillation of the hilltop axionic curvaton discussed later on can be
treated simply as sinusoidal, since the curvaton quickly settles
down to the part of its potential that is well approximated by a quadratic one. 
However, its cosine type potential~(\ref{eq:sin}) which is flatter than a
quadratic may allow formation of oscillating inflaton condensates
\cite{Bogolyubsky:1976yu,Bogolyubsky:1976nx,Gleiser:1993pt,Copeland:1995fq,Kasuya:2002zs}
during the initial oscillations.
Here we remark that such oscillons, even if they formed, are not
expected to alter the
above analyses since their energy density redshifts as nonrelativistic matter,
and also because their dynamics should not affect perturbations at the
CMB scales that are super-horizon by the time the oscillons form.} 
Then the curvaton energy density redshifts similarly to
nonrelativistic matter after the onset of the oscillations until 
the curvaton decays into radiation.
On the other hand, we consider the inflaton to
behave as matter from the end of inflation until reheating when it
decays into radiation.  The energy 
density of the curvaton before the beginning of its oscillation is
assumed to be negligibly tiny compared to the total energy of the
Universe, having little effect on the expansion history.  

Supposing the curvaton field fluctuations to be nearly Gaussian with 
$\mathcal{P}_{\delta \sigma} (k) = (H|_{k = aH } / 2 \pi)^2$ at the time when the
comoving wave mode~$k$ exits the horizon, then using the $\delta \mathcal{N}$-formalism
\cite{Starobinsky:1986fxa,Sasaki:1995aw,Wands:2000dp,Lyth:2004gb}, the
power spectrum of the density perturbations at the CMB scale is expressed as~\cite{Kawasaki:2011pd}
\begin{equation}
 \mathcal{P}_\zeta = \left(\frac{\partial \mathcal{N}}{\partial \sigma_*} 
  \frac{H_*}{2 \pi} \right)^2,
\label{eq-ps}
\end{equation}
with 
\begin{equation}
 \frac{\partial \mathcal{N}}{\partial \sigma_*} = \frac{r}{4+3 r}
  \left(1 - X(\sigma_{\mathrm{osc}})\right)^{-1}
 \left\{\frac{V'(\sigma_{\mathrm{osc}})}{V(\sigma_{\mathrm{osc}})} -
  \frac{3 X(\sigma_{\mathrm{osc}})}{\sigma_{\mathrm{osc}}}  \right\}
 \frac{V'(\sigma_{\mathrm{osc}})}{V'(\sigma_*)}.
 \label{NfuncX}
\end{equation}
Here, the subscript~$*$ denotes values when the CMB scale exits the
horizon, and $r$ is the energy density ratio between the curvaton and
radiation (which originates from the inflaton) upon curvaton decay
\begin{equation}
 r \equiv \left. \frac{\rho_\sigma }{\rho_r} \right|_{\mathrm{dec}}.
\label{def_r}
\end{equation}
The function~$X$ denotes effects due to the non-uniform onset of
the curvaton oscillations (which are absent for a purely quadratic
curvaton potential), defined as follows:
\begin{equation}
 X(\sigma_{\mathrm{osc}}) \equiv  
 \frac{1}{2(c-3)} \left(\frac{\sigma_{\mathrm{osc}}
	  V''(\sigma_{\mathrm{osc}})}{V'(\sigma_{\mathrm{osc}})} - 1
		  \right),
\label{Xdef}
\end{equation}
where the constant~$c$ is given in (\ref{ciroiro}).

From the above expressions, the spectral index of the
linear order perturbations follows as
(note that the scale-dependence of (\ref{NfuncX}) shows up only
through~$\sigma_*$, since $\sigma_{\mathrm{osc}}$ and $r$ are independent
of the comoving wave number)
\begin{equation}
 n_s -1 \equiv \frac{d}{d \ln k}\ln \mathcal{P}_{\zeta} =
 2\frac{\dot{H}_*}{H_*^2} + \frac{2}{3} \frac{V''(\sigma_*)}{H_*^2}.
\label{eq:ns}
\end{equation}
The recent observations strongly suggest that the density perturbation power
spectrum is red-tilted,  $n_s = 0.968 \pm 0.012$~\cite{Komatsu:2010fb}. 
This requires that the curvaton potential  be tachyonic 
and the size of the curvature must  be of order $10$\,\% of the Hubble
parameter during inflation, unless the inflaton is allowed to take
super-Planckian field values, or some special configurations are arranged
in the inflationary setup (cf. Footnote~\ref{foot1}.)

Curvatons also generate local-type\footnote{Strictly speaking, 
bispectra from curvatons have shapes similar to, but may not exactly be of the ``local
form''~\cite{Komatsu:2001rj}, especially when $f_{\mathrm{NL}}$ is strongly
scale-dependent~\cite{Byrnes:2010xd,Byrnes:2011gh,Kobayashi:2012ba}.
However we note that for axionic curvatons with sinusoidal potentials,
the running of~$f_{\mathrm{NL}}$ is tied to the running
of the spectral index, and thus 
strictly constrained to be small by current observations~\cite{Kobayashi:2012ba}.}
bispectrum, whose amplitude is 
represented by the non-linearity parameter~$f_{\mathrm{NL}}$. This is
given by
\begin{equation}\label{fNLfuncX}
\begin{split}
 f_{\mathrm{NL}}  & =
 \frac{5}{6} \frac{\partial^2 \mathcal{N}}{\partial
  \sigma_*^2} 
 \left( \frac{\partial \mathcal{N}}{\partial \sigma_*} \right)^{-2} \\
 & =
  \frac{40 (1+r)}{3 r (4+3 r)}  
 + \frac{5 (4+3 r)}{6 r}
 \left\{\frac{V'(\sigma_{\mathrm{osc}})}{V(\sigma_{\mathrm{osc}})}-
 \frac{3 X(\sigma_{\mathrm{osc}})}{\sigma_{\mathrm{osc}}}  \right\}^{-1}  
  \Biggl[(1-X(\sigma_{\mathrm{osc}}))^{-1} X'(\sigma_{\mathrm{osc}})    \\
 & \qquad 
  + \left\{\frac{V'(\sigma_{\mathrm{osc}})}{V(\sigma_{\mathrm{osc}})} -
 \frac{3X(\sigma_{\mathrm{osc}})}{\sigma_{\mathrm{osc}}}  \right\}^{-1} 
 \left\{\frac{V''(\sigma_{\mathrm{osc}})}{V(\sigma_{\mathrm{osc}})} -
 \frac{V'(\sigma_{\mathrm{osc}})^2}{V(\sigma_{\mathrm{osc}})^2} -
 \frac{3 X'(\sigma_{\mathrm{osc}})}{\sigma_{\mathrm{osc}}} + \frac{3
 X(\sigma_{\mathrm{osc}})}{\sigma_{\mathrm{osc}}^2} 
 \right\} \\
 & \qquad \qquad \qquad \qquad \qquad \qquad \qquad \qquad \qquad \qquad 
 \, \, 
  + \frac{V''(\sigma_{\mathrm{osc}})}{V'(\sigma_{\mathrm{osc}})} -
 (1-X(\sigma_{\mathrm{osc}}))
 \frac{V''(\sigma_*)}{V'(\sigma_{\mathrm{osc}})} 
\Biggr].
\end{split}
\end{equation}
A quadratic potential $V \propto \sigma^2$ realizing $X(\sigma_{\rm osc})
= 0$ reproduces the known result for quadratic curvatons whose
$f_{\mathrm{NL}}$ is determined only by~$r$.

Let us also rewrite the energy density ratio~$r$ (\ref{def_r}) 
in terms of the inflaton and curvaton parameters:
\begin{equation}\label{anar}
 r  = \mathrm{Max.}\left[
\frac{V(\sigma_{\mathrm{osc}})}{3 M_p^2 H_{\mathrm{osc}}^{3/2}
 \Gamma_\sigma^{1/2}} \times \mathrm{Min.}\left(
1, \,  \frac{\Gamma_\phi^{1/2}}{H_{\mathrm{osc}}^{1/2}} 
\right),
\left\{
\frac{V(\sigma_{\mathrm{osc}})}{3 M_p^2 H_{\mathrm{osc}}^{3/2}
 \Gamma_\sigma^{1/2}} \times \mathrm{Min.}
\left(
1, \,  \frac{\Gamma_\phi^{1/2}}{H_{\mathrm{osc}}^{1/2}} 
\right) 
\right\}^{4/3}
\right],
\end{equation}
where $M_p \simeq 2.4 \times  \GEV{18}$ is the reduced Planck mass, and 
the first and second terms in the Max. parentheses correspond to
the curvaton being subdominant and dominant at its decay,
respectively, while the Min. parentheses are due to whether the onset
of oscillation is after or before reheating.  $\Gamma_\phi$ and
$\Gamma_\sigma$ are constants that denote, respectively, the decay rates of
the inflaton and the curvaton. 
We note that in obtaining the above
results, we have adopted the sudden decay
approximation where the scalar fields suddenly decay into radiation
when $H = \Gamma$.

Finally, the curvaton field value at the onset of the
oscillations~$\sigma_{\mathrm{osc}}$ is obtained by
integrating~(\ref{CDapp}),
\begin{equation}
 \int^{\sigma_{\mathrm{osc}}}_{\sigma_*} \frac{d\sigma}{V'} =
 -\frac{\mathcal{N}_*}{3 H_{\mathrm{inf}}^2} - \frac{1}{2 c (c-3)
 H_{\mathrm{osc}}^2},  \label{sigmaosc}
\end{equation}
which can be solved for $\sigma_{\rm osc}$ as a function
of~$\sigma_*$.\footnote{When (\ref{sigmaosc}) admits as solutions for
$\sigma_{\rm osc}$ both positive and negative values, one should take
the sign of $\sigma_{\rm osc}$ to match with that of
$\sigma_*$.} Here, $\mathcal{N}_*$ is the number of e-folds during inflation
between the horizon exit of the CMB scale and the end of inflation,
$c$ is given in (\ref{ciroiro}), 
and $H_{\mathrm{inf}}$ is the inflationary Hubble scale (we are
assuming a nearly constant Hubble parameter during inflation, thus
$H_{\mathrm{inf}} \simeq H_*$). 

Therefore by combining the above expressions, one can compute the 
density perturbations from a curvaton with a generic
potential~$V(\sigma)$, given the curvaton field value at the CMB scale
horizon exit~$\sigma_*$, the decay rates of the inflaton~$\Gamma_\phi$
and curvaton~$\Gamma_\sigma$, the inflationary
scale~$H_{\mathrm{inf}}$, and the duration of
inflation~$\mathcal{N}_*$.

\subsection{Case Study: Hilltop Curvatons}
\label{subsec:hilltop}

As an example that will be relevant for analyzing axionic curvatons in the
next section, here let us apply the above generic results to 
a curvaton located at the hilltop, whose potential 
around $\sigma_{\mathrm{osc}}$ and $\sigma_*$ is well approximated by
\begin{equation}
 V(\sigma) = V_0 - \frac{1}{2} m^2 (\sigma - \sigma_0)^2,
\end{equation}
where $m$,
$\sigma_0$, and $V_0 (>0)$ are constants. 
Without loss of generality, we assume $0 <
\sigma_{\mathrm{osc}} < \sigma_* < \sigma_0$. 
Then one can check that when the curvaton is close enough to the hilltop
to satisfy
\begin{equation}
 \sigma_{\mathrm{osc}}  \gg \sigma_0 - \sigma_{\mathrm{osc}},
 \quad
 V_0  \gg m^2 ( \sigma_0 - \sigma_{\mathrm{osc}} )^2,
 \label{eq36}
\end{equation}
then the resulting power spectrum~(\ref{eq-ps}) and the
non-Gaussianity~(\ref{fNLfuncX}) take the form
\begin{equation}
 \mathcal{P}_\zeta^{1/2} \simeq \frac{3r}{4+3 r}\frac{\sigma_0 -
  \sigma_{\mathrm{osc}}}{\sigma_0 - \sigma_*} \frac{H_{*}}{2
  \pi \sigma_{\mathrm{osc}}},  \label{P}
\end{equation}
\begin{equation}
 f_{\mathrm{NL}} \simeq \frac{5 (4+3 r)}{18 r}
  \frac{\sigma_{\mathrm{osc}}}{\sigma_0 - \sigma_{\mathrm{osc}}},  \label{fNL}
\end{equation}
with spectral index~(\ref{eq:ns})
\begin{equation}
 n_s -1 = 2 \frac{\dot{H}_*}{H_*^2 }
-\frac{2}{3} \frac{m^2}{H_{*}^2} .
\label{spindex}
\end{equation}
The equation~(\ref{sigmaosc}) which relates $\sigma_*$ and
$\sigma_{\mathrm{osc}}$ gives 
\begin{equation}
 \ln \left( \frac{\sigma_0 - \sigma_*}{\sigma_0 - \sigma_{\mathrm{osc}}} \right)
 \simeq 
 - \frac{1}{2 (c-3) } \frac{\sigma_{\mathrm{osc}}}{\sigma_0
       - \sigma_{\mathrm{osc}}} ,
 \label{eq311}
\end{equation}
where we dropped the $H_{\mathrm{inf}}^2$ contribution
on the right hand side from the condition~(\ref{eq36}) and also by
assuming $m^2 / H_{\mathrm{inf}}^2 \lesssim 10^{-2}$.
As the initial value~$\sigma_*$ is shifted towards the hilltop, 
$\sigma_{\mathrm{osc}}$ approaches $\sigma_0$ much slower than
$\sigma_*$ does since the left hand side is logarithmic.
Therefore as one approaches the hilltop, $\mathcal{P}_\zeta$ (\ref{P}) blows
up due to the enhancement factor $(\sigma_0 - \sigma_{\mathrm{osc}}) /
(\sigma_0 - \sigma_*)$, while $f_{\mathrm{NL}}$ (\ref{fNL}) increases
slowly. We also note that the value of $f_{\mathrm{NL}}$ is greater than
one even when $r \gg 1$, from~(\ref{eq36}). 
The extreme amplification of the linear perturbations
corresponds to the curvaton taking longer time to
start its oscillation when starting closer to the hilltop. 

Before ending this section, we should remark that in the extreme hilltop limit,
the approximation~(\ref{CDapp}) for the curvaton dynamics mildly breaks
down before the curvaton starts to oscillate. This gives rise to
errors of $\mathcal{O}(1)$ for the above results in this limit.
However, the above analytic 
expressions suffice for our order of magnitude estimations on
axionic curvatons in the next section. We will also
carry out numerical computations when further accuracy is required,
e.g., when calculating predictions on~$f_{\mathrm{NL}}$.

\section{Axionic Curvatons}
\label{sec:pNGcurvaton}

Now let us move on to the investigation of axionic
curvatons, which is the main topic of this paper. As was explained in
the introduction, we focus on the case where the curvaton is a
pseudo-Nambu-Goldstone boson of a broken U(1) symmetry, possessing a
periodic potential of the form 
\begin{equation}
 V(\sigma) = \Lambda^4 \left[ 1 - \cos \left(\frac{\sigma}{f} \right)
		       \right] ,
\label{eq:sin}
\end{equation}
where $f$ and $\Lambda $ are mass scales. Without loss of generality, we
restrict the initial field value to lie within the range $0 < \sigma_* < \pi f$. 
The curvaton's effective mass at the potential minimum is denoted by
\begin{equation}
 m_\sigma = \frac{\Lambda^2}{f} .
\label{msigma}
\end{equation}
Then supposing that the coupling of the axionic curvaton with its decay
product is suppressed by the symmetry breaking scale~$f$, the curvaton
decay rate takes the value
\begin{equation}
 \Gamma_\sigma = \frac{\beta }{16 \pi} \frac{m_\sigma^3}{f^2} = 
 \frac{\beta }{16 \pi } \frac{\Lambda^6}{ f^5 },
\label{decayrate}
\end{equation}
where the constant $\beta$ is naively of order unity.
In the following, we ignore the time-variation of the Hubble parameter during
inflation, and especially, neglect the $\dot{H}$ contribution to the
spectral index (\ref{eq:ns}).
In other words, we do not consider inflationary
models with rather large $|\dot{H}/ H^2|$ which requires 
super-Planckian field ranges or some special
configurations.\footnote{Assuming single-field canonical slow-roll 
inflation, the Lyth bound~\cite{Lyth:1996im} relates the time-variation
of the Hubble parameter with the inflaton field~$\phi$ range as 
\begin{equation*}
 \frac{1}{M_p^2} \left(\frac{d\phi}{d\mathcal{N}} \right)^2 
\simeq - 2 \frac{\dot{H}}{H^2},
\end{equation*}
where $M_p$ is the reduced Planck mass and $\mathcal{N}$ the e-folding number.
Thus $|\dot{H} / H^2 |$ as large as to give sizable contribution to the
spectral index~(\ref{eq:ns}) whose typical value is $n_s  \approx 0.968$ 
(WMAP central value)
normally requires a super-Planckian field range for the inflaton.
The field range bound may be alleviated by inflaton potentials giving
sudden changes to $d \phi / d \mathcal{N}$ during
inflation~\cite{BenDayan:2009kv}.\label{foot1}}
Hence the axionic curvaton need to be located beyond the inflection point
during inflation, i.e. $0.5 < \sigma_*/ \pi f < 1 $, in order to source
a red-tilted power spectrum. 

The axionic curvaton with $\sigma_*  \ll \pi f $ whose
potential is well approximated by a quadratic was studied
in~\cite{Dimopoulos:2003az}, and the whole potential including the
hilltop region was investigated in~\cite{Kawasaki:2011pd}.
There it was shown along the line of discussion in
Section~\ref{subsec:hilltop}, that unless the axionic curvaton is initially
located close to the hilltop, both the inflation and reheating scales
need to be very high. For e.g., for $\sigma_* / \pi f =
0.75$ to satisfy both the WMAP normalization $P_{\zeta} \approx 2.42
\times 10^{-9}$ and the spectral index $n_s
\approx 0.96$, then $H_{\mathrm{inf}} \gtrsim 10^{13} \, \mathrm{GeV} $ and
$\rho_{\mathrm{reh}}^{1/4} \gtrsim 10^{13} \, \mathrm{GeV}$ are
required,  where $\rho_{\mathrm{reh}}$ represents the radiation energy density
at the reheating.
This is because the spectral index of order $1 - n_s \sim
0.01$ requires a rather large curvaton mass $m_{\sigma} \sim 0.1
H_{\mathrm{inf}}$, forcing the curvaton to start its oscillation soon
after the end of inflation. Hence without high inflation and reheating
scales, the curvaton cannot even come close to dominating the 
Universe to source measurable density perturbations.\footnote{The
curvaton's effective mass during inflation is decoupled from the mass
at the potential minimum~(\ref{msigma}) when the curvaton is close to
the inflection point, 
i.e. $\sigma_* / \pi f  \approx 0.5$, however in such case even higher 
inflation/reheating scales are required.} 
The story is quite different for an axionic curvaton in the hilltop region,
where the onset of the oscillation is delayed and curvaton
domination is allowed with lower inflation/reheating scales. This, together
with the amplification of the linear perturbations in the hilltop limit
(cf. discussions around~(\ref{eq311})), makes axionic curvatons compatible
with many orders of magnitude of the inflation and reheating scales. 

In light of the above considerations, in this section we elaborate on
axionic curvatons in the hilltop region, which dominate the Universe before
decaying into radiation. We will find that this particular limit of axionic
curvatons has interesting predictions, especially in terms of the
non-Gaussianity. 

\subsection{Parameter Space in the Hilltop Regime}

The axionic curvaton model has five free parameters, which are the symmetry
breaking scale~$f$, the effective mass~$m_\sigma = \Lambda^2 / f$, the
curvaton field value at CMB scale horizon exit~$\sigma_*$, the
inflationary scale~$H_{\mathrm{inf}}$, and the inflaton decay
rate~$\Gamma_\sigma$. However, since we are focusing on a curvaton that
dominates the Universe before it decays, 
as long as there exists a parameter window which allows $ r \gg 1$, 
the cosmological observables do not depend on the explicit value
of~$r$ or $\Gamma_\sigma$. In this sense, the dominant axionic curvaton is
actually a four parameter model. 

Strictly speaking, there are three more parameters: the e-folding
number~$\mathcal{N}_*$ between the CMB scale horizon exit and the end of 
inflation, the constant~$c$ (\ref{ciroiro}) representing
whether $t_{\mathrm{reh}} \gtrless t_{\mathrm{osc}}$ (though this is 
determined when the other parameters such as~$\Gamma_\phi$ are fully
given), and $\beta$ in 
(\ref{decayrate}) parameterizing the 
curvaton decay rate. $\mathcal{N}_*$ determines how much the curvaton
rolls during inflation (cf. (\ref{sigmaosc})), however such rolling
is negligible compared to that in the post-inflationary
era as seen in (\ref{eq311}), and thus has little effects on the
model. Hence we simply fix the e-folding number to $\mathcal{N}_* = 50$
in the following discussions. As for~$c$,
whether reheating happens before/after the onset of the curvaton
oscillations do not affect the allowed parameter window for $f$ and
$m_\sigma$, but give slightly different predictions
on~$f_{\mathrm{NL}}$. This will be discussed in Section~\ref{subsec:OP}. 
The parameter~$\beta$ for the decay rate is set to unity in the
following, and implications of $\beta$ taking other values are also
discussed later. 

\vspace{\baselineskip}

\begin{figure}[p]
 \begin{minipage}{.48\linewidth}
  \begin{center}
 \includegraphics[width=\linewidth]{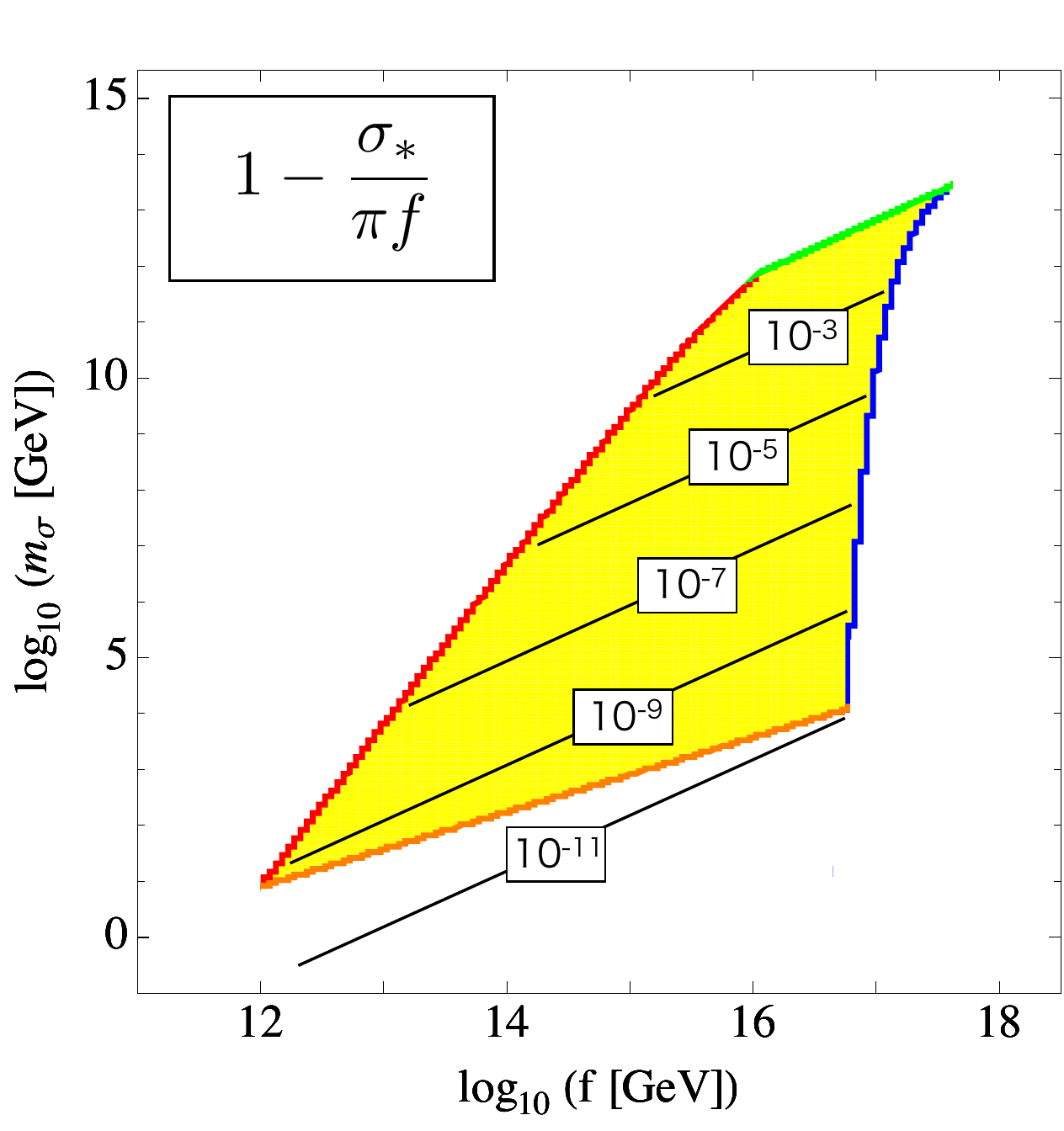}
  \end{center}
 \end{minipage} 
 \begin{minipage}{0.01\linewidth} 
  \begin{center}
  \end{center}
 \end{minipage} 
 \begin{minipage}{.48\linewidth}
  \begin{center}
 \includegraphics[width=\linewidth]{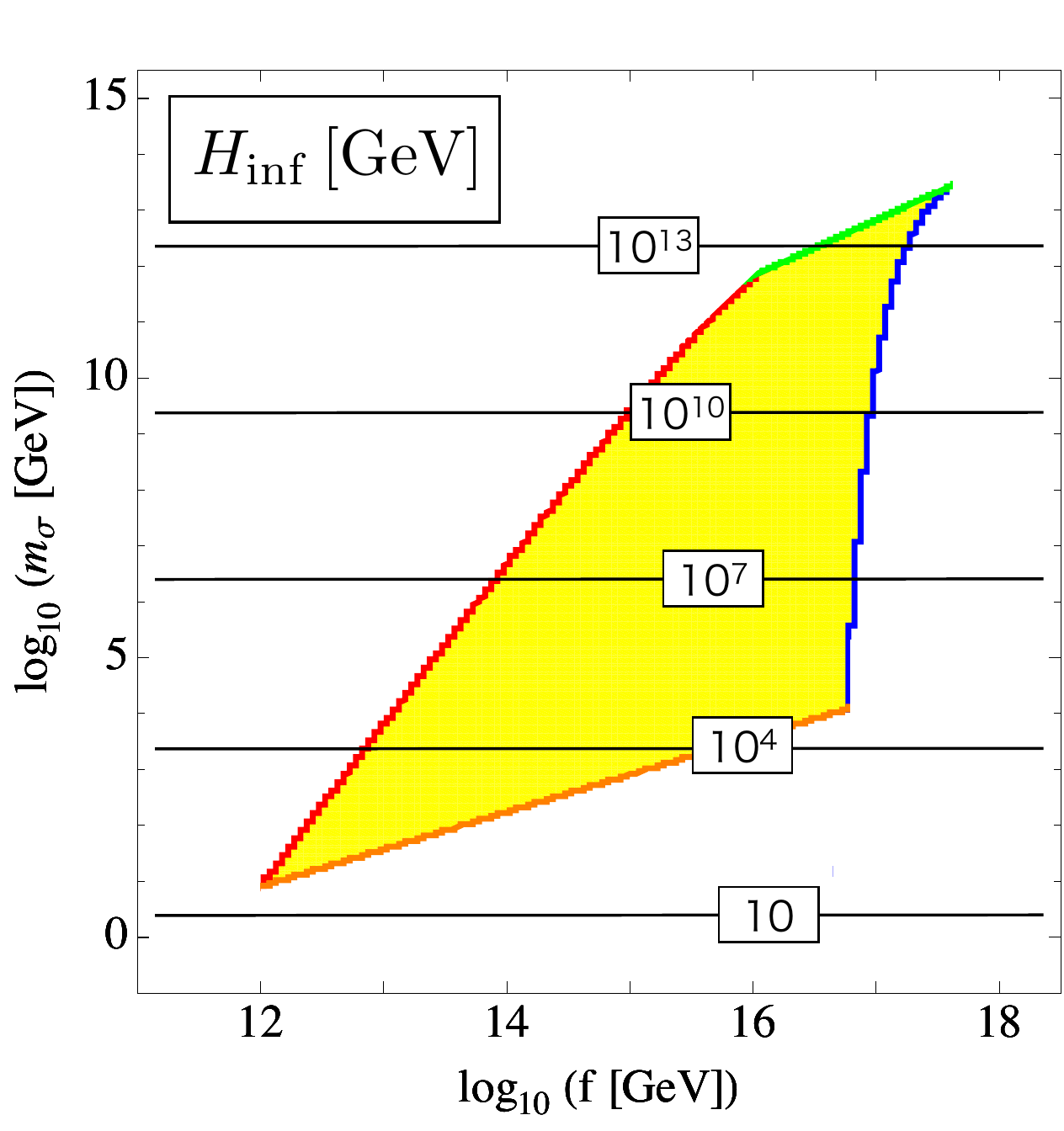}
  \end{center}
 \end{minipage} 
 \begin{minipage}{.48\linewidth}
  \begin{center}
 \includegraphics[width=\linewidth]{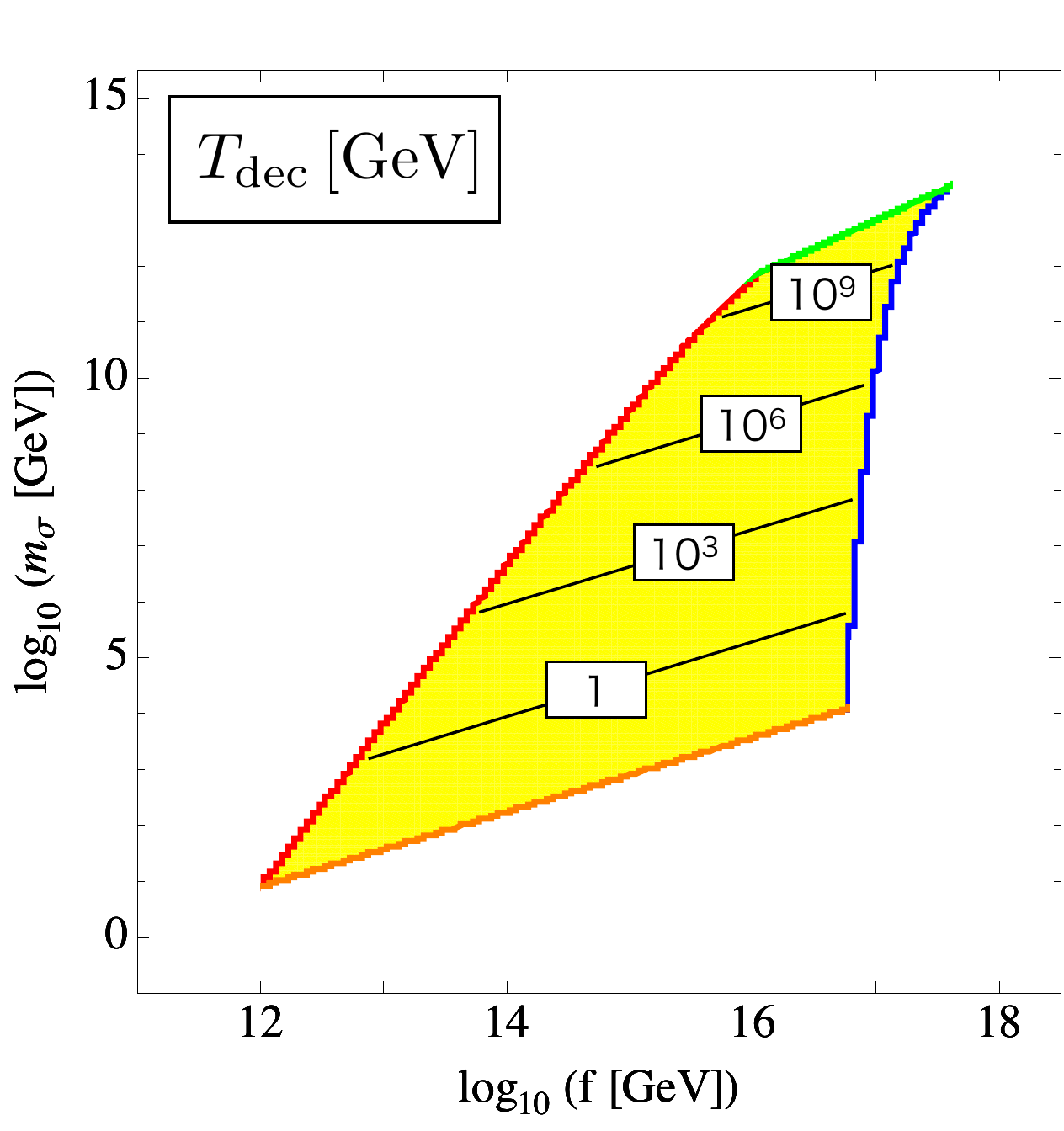}
  \end{center}
 \end{minipage} 
 \begin{minipage}{0.01\linewidth} 
  \begin{center}
  \end{center}
 \end{minipage} 
 \begin{minipage}{.48\linewidth}
  \begin{center}
 \includegraphics[width=\linewidth]{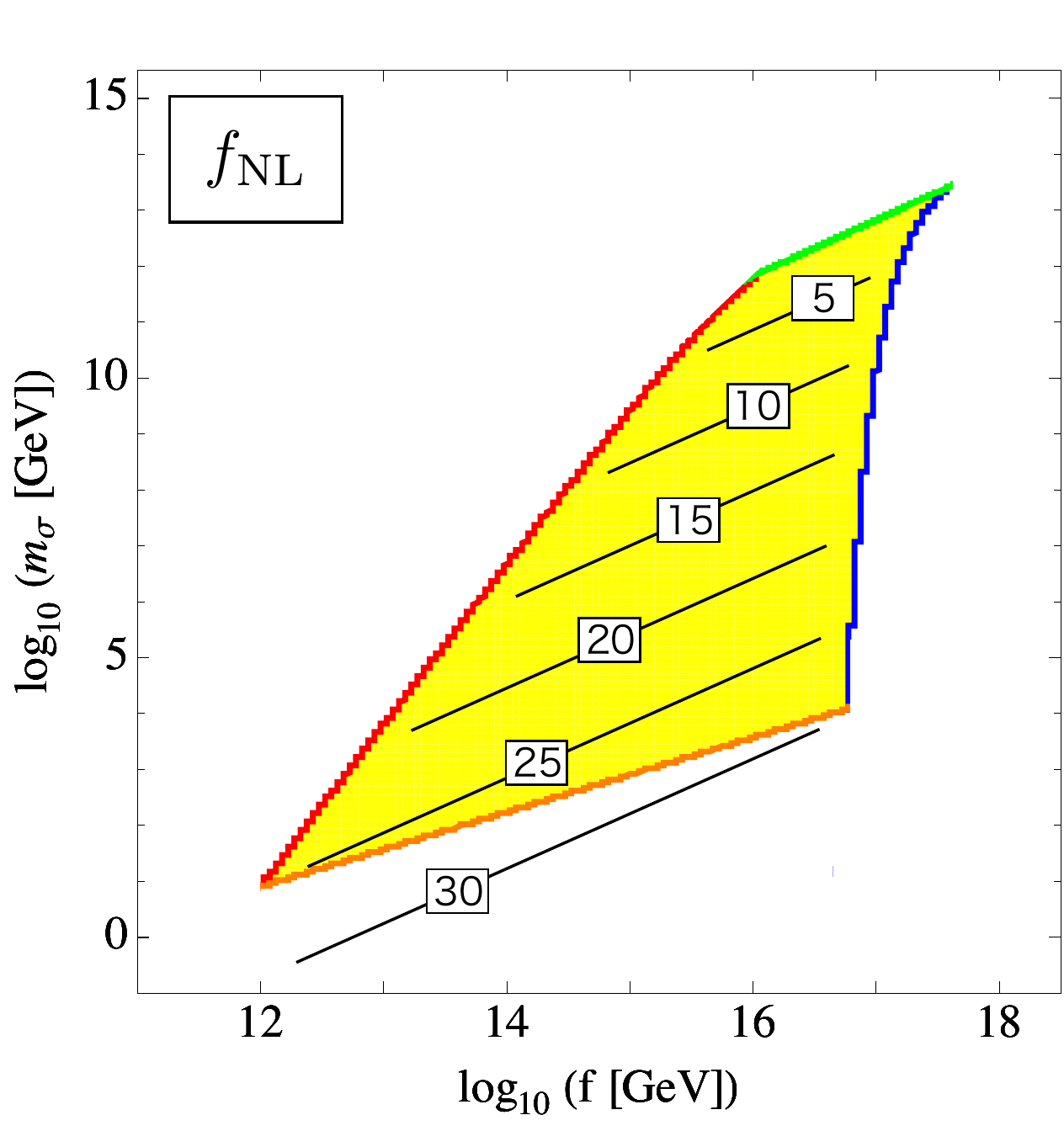}
  \end{center}
 \end{minipage} 
  \caption{Parameter space for a dominant axionic curvaton in the hilltop. The allowed
 window is shown as the yellow region, which is bordered by the hilltop
 condition~(\ref{green}) (upper green boundary), the requirement that
 the curvaton does not inflate the Universe~(\ref{blue}) (right blue), BBN
 constraint~(\ref{orange}) (lower orange), and the dominant
 condition~(\ref{red}) (left red). These conditions have been adopted in
 order to study the peculiar behavior of the hilltop curvaton. However the curvaton
 mechanism can still work when relaxing some of them, see the discussions
 in Appendix~\ref{app:infcurv} and footnote~\ref{foot:10}. 
 The contour lines on each figure denote the following quantities. Upper
 left: The curvaton value at CMB scale horizon exit $(\pi f - \sigma_*)
 / \pi f$. Upper right: Inflationary scale~$H_{\mathrm{inf}}$ in units
 of GeV. Lower left: Decay temperature~$T_\mathrm{dec}$ of the curvaton
 in units of GeV. Lower right: Non-Gaussianity~$f_{\mathrm{NL}}$ for the case of 
 $t_{\mathrm{osc}} < t_{\mathrm{reh}}$. The values of  $f_{\mathrm{NL}}$
 slightly increases when $t_{\mathrm{osc}} > t_{\mathrm{reh}}$.}
  \label{fig:hilltoppNG}
\end{figure}

Out of the four parameters, $H_{\mathrm{inf}}$ and $\sigma_* / f
$ can be fixed from the WMAP normalization 
\begin{equation}
 \mathcal{P}_\zeta \approx 2.4 \times 10^{-9},
\label{COBE}
\end{equation}
and also from requiring the spectral index to be consistent with the WMAP
bound 
\begin{equation}
 n_s \approx 0.96. 
\label{0.96}
\end{equation}
Later on we will see that the detailed value of the
spectral index, as long it is not so close to unity, only have minor
effects on axionic curvatons. 
Hence we are left with two parameters for the axionic curvaton $f$ and
$m_\sigma$. Order of magnitude constraints on these parameters can be
obtained using the analytic formulae in Section~\ref{sec:review} (or in
Section~\ref{subsec:hilltop}), which we present in
Figure~\ref{fig:hilltoppNG}. The yellow region corresponds to the
allowed window for a dominant axionic curvaton in the hilltop.
$H_{\mathrm{inf}}$ and $\sigma_* / f$ are fixed to appropriate values
by the observational constraints (\ref{COBE}) and (\ref{0.96}) at each
point in the window, as indicated in the upper figures showing their
contour lines. This also fixes the curvaton decay rate via
(\ref{decayrate}), cf. lower left figure. (The relativistic degrees of
freedom is fixed to $g_*  = 100$ upon drawing the $T_\mathrm{dec}$ contours).
On the other hand, the inflaton decay
rate~$\Gamma_\phi$ is not fixed at each point but is allowed to take
values within a certain range, as we will soon explain. 
The lower right figure shows contour lines for the
non-Gaussianity~$f_{\mathrm{NL}}$, which is typically a few tens.
Here we note that since the analytic formulae in
the previous section can contain $\mathcal{O}(1)$ errors
(cf. discussions at the end of Section~\ref{subsec:hilltop}), the
$f_{\mathrm{NL}}$ values have been computed numerically.
We have shown the $f_{\mathrm{NL}}$ contours inside the allowed window
where the constraints described in the following are well satisfied, but
the values can be modified at regions very close to the boundaries.

We find that the allowed window is constrained by the following four conditions: 
The upper edge (green line) is set by the requirement that the curvaton
initially lies in the hilltop regime,
\begin{equation}
 \frac{\sigma_*}{\pi f} > 0.9.
\label{green}
\end{equation}
Recall that a non-hilltop axionic curvaton can work only with very high
inflation/reheating scales. The right edge (blue line) denotes the
requirement that the curvaton be 
subdominant until it starts its oscillation, 
\begin{equation}
 V(\sigma_{\mathrm{osc}}) < 0.1 \times 3 M_p^2 H_{\mathrm{osc}}^2.
\label{blue}
\end{equation}
When going beyond this boundary, the curvaton starts to drive a
secondary inflation. 
A rather strict relationship between $H_{\mathrm{in}}$ and $\sigma_* /
f$ is required for such inflating curvatons to work, as is discussed in
Appendix~\ref{app:infcurv}. The lower edge (orange line) requires the
curvaton to decay at temperatures higher than 5 MeV in order not to ruin
Big Bang Nucleosynthesis (BBN) 
\cite{Kawasaki:1999na,Kawasaki:2000en,Hannestad:2004px,Ichikawa:2005vw},
i.e.
\begin{equation}
 3 M_p^2 \Gamma_{\sigma}^2 > \frac{\pi^2}{30} g_* (5\,
  \mathrm{MeV})^4 ,
\label{orange}
\end{equation}
with the relativistic degrees of freedom $g_* = 10.75$. 
Finally, the left edge (red line) follows from the dominant condition~\footnote{
If the curvaton is subdominant at the decay, namely, $r \ll 1$, the non-Gaussianity
tends to be too large, and one has to tune the decay rates of the curvaton and the inflaton
so that $r \gtrsim 0.01$ is realized in order to be consistent with the observations. 
We have nothing new to add to this possibility in our context.\label{foot:10}
}
\begin{equation}
 r > 10.
\label{red}
\end{equation}
These conditions give the most stringent constraints on the hilltop axionic
curvaton, and other requirements for a consistent curvaton scenario are
satisfied in the window bordered by (\ref{green}) - (\ref{red}). 

Let us also lay out such satisfied conditions:
Firstly, the curvaton energy density is negligibly small
during inflation, 
\begin{equation}
 V(\sigma_*) \ll 3 M_p^2 H_{\mathrm{inf}}^2.
 \label{cond3}
\end{equation}
Moreover, quantum fluctuations during inflation should not make the curvaton
jump over its potential minimum in order to avoid the resulting density
perturbations from being highly non-Gaussian, or over the maximum to
avoid domain walls,
\begin{equation}
 \frac{H_{\mathrm{inf}}}{2 \pi } \ll \sigma_*
 \ll \pi f -  \frac{H_{\mathrm{inf}}}{2 \pi }.
\label{nodomain}
\end{equation}
In the hilltop region, the classical rolling becomes suppressed, which can
compete with the quantum fluctuations during inflation.\footnote{The
randomized case of axionic
curvatons at the potential minimum is discussed in~\cite{Dimopoulos:2003az}.} 
The curvaton's classical rolling dominates over the quantum fluctuations if 
\begin{equation}
  \frac{3}{2 \pi} \frac{H_{\mathrm{inf}}^3}{V'(\sigma_*)} 
  \ll 1,
\label{3.12}
\end{equation}
where the curvaton is considered to slow-roll due to (\ref{cond3}) and 
the lightness condition that follows from the spectral
index~(\ref{0.96}). 
Furthermore, the curvaton
decay should happen after reheating and the onset of the oscillations,
\begin{equation}
 \Gamma_\sigma < \Gamma_\phi,\, H_{\mathrm{osc}}.
\label{3.13}
\end{equation}
The mass~$m_\sigma$ is required to be larger than the curvaton
decay temperature, in order to avoid possible backreaction effects
to the curvaton's perturbative decay (see
e.g. \cite{Kolb:2003ke,Yokoyama:2005dv,Drewes:2010pf}). Assuming instant
thermalization, this condition is written roughly as  
\begin{equation}
 m_\sigma^2 > (3 M_p^2 \Gamma_\sigma^2)^{1/2}.
\end{equation}
As for the inflaton sector, the energy scale of reheating (= inflaton decay) is
lower than that of inflation, while an upper bound on the inflationary 
scale is given by constraints on primordial gravitational waves.
The 7-year WMAP+BAO+$H_0$ gives $\mathcal{P}_T / \mathcal{P}_\zeta <
0.24$ (95\% CL), which translates into\footnote{When the inflation scale
is high enough to saturate the bound~(\ref{cond14}), then depending on the
inflationary mechanism, one can expect to have a contribution to the
spectral index from a non-vanishing $\dot{H}/H^2$ (see also
Footnote~\ref{foot1}), as well as the central value of the spectral
index bounds~(\ref{0.96}) being slightly shifted.
We ignore such effects for axionic curvatons, since they can modify the
results only at the vicinity of the upper right corner of the allowed window
in Figure~\ref{fig:hilltoppNG}.}
\begin{equation}
 \Gamma_\phi < H_{\mathrm{inf}} < 1.3 \times 10^{14}\,
  \mathrm{GeV}. \label{cond14} 
\end{equation}
Let us repeat that all the requirements (\ref{cond3}) - (\ref{cond14})
are satisfied in the allowed window of Figure~\ref{fig:hilltoppNG}. 

\vspace{\baselineskip}

We should also remark on the constraints on the reheating scale~$\Gamma_\phi$
before ending this subsection. As we have noted above, dominant axionic
curvatons are insensitive to the
explicit value of~$\Gamma_\phi$. The only constraints on~$\Gamma_\phi$
are that the inflaton should decay after the end of inflation
(\ref{cond14}) but before the curvaton decay (\ref{3.13}). 
For the case of $t_{\mathrm{osc}} < t_{\mathrm{reh}}$, 
the dominant condition~(\ref{red}) sets an additional lower bound
on~$\Gamma_\phi$, cf.~(\ref{anar}).\footnote{For the case of
$t_{\mathrm{osc}} < t_{\mathrm{reh}}$, 
the left red edge actually denotes where
the upper bound on~$\Gamma_\phi$ set by $ \Gamma_\phi<
H_{\mathrm{osc}}$ and 
the lower bound from the dominant condition~(\ref{red}) take the same
values. In other words, 
$t_{\mathrm{osc}} < t_{\mathrm{reh}}$ and (\ref{red}) are incompatible
beyond the red line. 
On the other hand, for the $t_{\mathrm{osc}} < t_{\mathrm{reh}}$ case, 
$r$ is independent of~$\Gamma_\phi$, which allows one to set the dominant
condition~(\ref{red}) independently of~$\Gamma_\phi$. 
However we note that (\ref{red}) produces the left red edge at the same place
on the $f-m_\sigma$ plane for both $t_{\mathrm{reh}} \gtrless
t_{\mathrm{osc}}$ cases.} 
The inflaton decay rate should take values within
these bounds at each point of the allowed window in
Figure~\ref{fig:hilltoppNG}. 
We note that the contour lines of various quantities in the figures are
obtained assuming that the values of~$\Gamma_\phi$ at each point do not
saturate the lower/upper bounds 
set by the above requirements. If, for example,  $\Gamma_\phi$ takes
lowest possible values saturating the dominant condition~(\ref{red})
for the $t_{\mathrm{osc}} < t_{\mathrm{reh}}$ case,
then $r$ becomes as small as $\approx 10$, slightly
modifying~$f_{\mathrm{NL}}$ from the shown values.

\subsection{Dependence on Various Parameters}
\label{subsec:OP}

The spectral index $n_s - 1 = -2 m_\sigma^2 / 3
H_{\mathrm{inf}}^2$ fixes the inflationary scale~$H_{\mathrm{inf}}$ 
proportional to the curvaton mass~$m_\sigma$, as is shown in the upper
right figure. The rather wide range allowed for~$m_\sigma$ is translated into
the axionic curvaton being compatible with inflationary scales with many
orders of magnitude. 

The upper left figure shows the curvaton's initial 
value~$\sigma_*$, which comes closer to the hilltop~$\pi
f $ towards the lower right corner of the allowed region. 
However when so close to the hilltop such that
$\sigma_* / \pi f \gtrsim 1-10^{-11}$, then the axionic curvaton either
ruins BBN, or drives a secondary inflation. For dominant curvatons $r
\gg 1$, the non-Gaussianity~$f_{\mathrm{NL}}$ is determined by how close
the curvaton initially is to the hilltop (cf. (\ref{fNL}) and
(\ref{eq311})), thus the $f_{\mathrm{NL}}$ contours run parallel to
those of $(\pi f - \sigma_*) / \pi f$. 
Here, recall that when shifting $\sigma_*$ towards the hilltop,
$\sigma_{\mathrm{osc}}$ increases much slower than $\sigma_*$ does. This
leads to a mild 
increase of $f_{\mathrm{NL}}$, whose largest possible value is
$ \sim 30$ for the case of $t_{\mathrm{osc}} < t_{\mathrm{reh}}$, and 
$ \sim 40$ for $t_{\mathrm{osc}} > t_{\mathrm{reh}}$. 
We also note that the regime beyond the right blue edge corresponds to axionic
curvatons driving a second inflationary stage. However, such inflating
axionic curvatons produce rather small non-Gaussianity, as discussed in
detail in Appendix~\ref{app:infcurv}. 

Now let us discuss the model dependence on other parameters.

\subsubsection*{Reheating Before/During Curvaton Oscillations}

Whether reheating happens before or during
the curvaton oscillation only slightly modify the curvaton velocity prior to
the oscillation. This does not affect our order of magnitude estimation
on the allowed window in the $f-m_\sigma$ plane, except for that 
the case of $t_{\mathrm{reh}} < t_{\mathrm{osc}} $ restricts the
inflaton decay rate~$\Gamma_\phi$ to lie within a rather narrow
range~\cite{Kawasaki:2011pd}. The $\sigma_* /
\pi f$, $H_{\mathrm{inf}}$, and $T_{\mathrm{dec}}$  contours are
also nearly the same for the two cases, however we 
note that $f_{\mathrm{NL}}$ can be slightly larger when
$t_{\mathrm{reh}} < t_{\mathrm{osc}} $. This is because a radiation
dominated Universe allows $\sigma$ to roll less compared to when
dominated by matter, and thus slightly makes $\sigma_{\mathrm{osc}}$
closer to the hilltop. The lower right figure shows the $f_{\mathrm{NL}}$
contours for $t_{\mathrm{reh}} > t_{\mathrm{osc}} $, 
but the case of $t_{\mathrm{reh}} < t_{\mathrm{osc}}$ increases the
$f_{\mathrm{NL}}$ values by up to $\sim 10$. 

\subsubsection*{Spectral Index}

The allowed window and non-Gaussianity are insensitive to the
explicit value of~$n_s$ (whether $n_s$ is, say, $0.94$ or $0.98$). 
However, if the spectral index is as close to
unity as $n_s > 0.99$, then the curvaton potential is required to be so
flat such that the quantum fluctuations can dominate over the curvaton's
classical rolling during inflation in the hilltop regime. 
When increasing $n_s$ beyond 0.99 towards unity, 
the condition~(\ref{3.12}) is violated first in the 
lower right corner of the allowed window in the $f$ - $m_\sigma$ plane, and
eventually in the entire window at $n_s \gtrsim 0.999$. 
Normally the model loses precise predictions if the quantum
fluctuations dominate over the classical rolling.
However we expect that the predictions for hilltop curvatons are
not affected much by such quantum jumps during inflation, since it is the
non-uniform onset of the curvaton oscillation that mainly generates the
linear and second order perturbations, and also because (\ref{nodomain})
is satisfied even for $n_s \approx 0.999$, i.e., the quantum jumps (even
when they dominate the curvaton dynamics) do not drastically
change the curvaton position during inflation.
We leave this question for future work, and let us close this paragraph
by stating that as long as $n_s \lesssim 0.99$, the detailed values of
the spectral index has little effect on axionic curvatons.

\subsubsection*{Curvaton Decay Rate}

We have been setting $\beta$ as unity in the curvaton decay
rate~(\ref{decayrate}). A further suppressed~$\Gamma_\sigma$ 
(i.e. smaller~$\beta$) delays the curvaton decay,
thus makes the BBN constraint~(\ref{orange}) more stringent, while
making it easier for the curvaton to dominate the Universe and
relaxes the dominant condition~(\ref{red}).
For example, $\beta = 10^{-3}$ tightens the lower edge (orange boundary) of
the allowed window in Figure~\ref{fig:hilltoppNG} by
$\Delta (\log_{10} m_\sigma)  \sim 1$, 
but pushes out the left edge (red) slightly (i.e. does not change the
order of~$f$).
It should be noted that the tightening of the BBN constraint results in
decreasing the largest possible value for~$f_{\mathrm{NL}}$, as can be
seen in the lower figure. 
For $\beta = 10^{-3}$, the maximum $f_{\mathrm{NL}}$ is about~$27$.

\section{Discussion}
\label{sec:Disc}

The upshot of our results is that an axionic curvaton generating density
perturbations consistent with current observations also generically
produce non-Gaussianity $f_{\rm NL}$ of ${\cal O}(10)$,
even when the curvaton dominates the Universe. In particular, 
as one can see from Fig.~\ref{fig:hilltoppNG},
$f_{\rm NL} = 20 \hyphen 40$ is realized for the curvaton mass
$m_\sigma = 10 \hyphen 10^6$\,GeV and the decay constant $f = 10^{12} \hyphen 10^{17}$\,GeV.
 
 What is the plausible candidate for the axionic curvaton?
Interestingly,  there are many moduli fields $(T)$ in the string theory, and they are massless
at the perturbative level because of the shift symmetry,
\bea
T &\rightarrow& T+ i \alpha,
\eea
where $\alpha$ is a real transformation parameter. After the moduli fields are stabilized
by non-perturbative effects and  SUSY breaking, 
the imaginary components of the moduli fields, namely the (string) axions, acquire a sinusoidal potential
like \EQ{eq:sin}.
The symmetry breaking scale $f$ is naively expected to be of order the GUT or Planck scale. Thus, the string axion is one of the plausible candidates for the axionic curvatons.\footnote{
The real component may play a role of the inflaton, in which case the moduli explains both the inflation and
the origin of density perturbations.} 

Recently, the standard-model like Higgs boson was discovered by the ATLAS and CMS 
experiments~\cite{:2012gk,:2012gu}. The observed Higgs boson mass is about 
$125 \,\mathchar`- \,126$\,GeV,  which can be explained if SUSY is realized at a relatively high scale~\cite{Okada:1990gg,Giudice:2011cg},
ranging from $10$\,TeV up to several tens PeV depending on the ratio of the up- and down-type Higgs boson VEVs.  
While the axion mass crucially depends on the stabilization mechanism, it is related to the gravitino mass
in a KKLT-type stabilization~\cite{Kachru:2003aw}, and so, it is conceivable that the axion mass is
not many orders of magnitude different from the suggested SUSY breaking scale in the visible sector.  It is intriguing that the axion with mass
of this order can generate a large non-Gaussianity within the reach of the Planck satellite.

The initial position of the  curvaton must be very close to the hilltop of the potential.
If some symmetries are restored at the maximum of the potential,
 the curvaton sits initially very close to the hilltop without any fine-tuning.
This is possible if one considers a moduli space spanned
by multiple scalar fields~\cite{Nakayama:2012gh}. 
To be concrete, we consider
a supersymmetric theory with the superpotential,
\begin{equation}
W\;=\; S(\mu^2 -  \chi^2 -  \phi^2).
\label{W}
\end{equation}
Here $S$, $\chi$ and $\phi$ are chiral superfields, and $\mu$ is a mass
scale that is real. We assume that both $\chi$ and $\phi$ parameterize D-flat directions
so that their origins are  enhanced symmetry points where the corresponding gauge fields
become massless.  In the supersymmetric
limit, there is a moduli space characterized by 
\begin{equation}
\chi^2 + \phi^2 = \mu^2,
\label{mod_space}
\end{equation}
where it should be noted that both $\chi$ and $\phi$ are complex scalar fields.
The scalar potential vanishes in the moduli space. 
There are two special symmetry-enhanced points, i.e., $\chi=0$ and $\phi = 0$.
The degrees of freedom orthogonal to the moduli space are heavy, and can be integrated out. For instance,
$\chi$ is heavy at $\phi \approx 0$, one of the symmetry enhanced points, 
and we can erase $\chi$ by using (\ref{mod_space}). In order to see that the potential has extrema
at those symmetry enhanced points,
let us introduce a soft SUSY breaking mass, $m^2 |\chi|^2$, which lifts the moduli space. 
 A similar soft SUSY breaking mass
can be introduced for $\phi$, but it does not change the argument. Since it is $\phi$ that is light at $\phi
\approx 0$, the effective potential can be written as
\begin{equation}
V_{\rm eff} \;=\; 
m^2 |\mu^2 - \varphi^2|,
\label{mod_space2}
\end{equation}
where we have supposed $m^2 > 0$ and minimized the angular component of $\phi$, and defined $\varphi \equiv
|\phi|$. Thus, $\phi=0$ is the local maximum. Note that one should write the effective
potential in terms of $\chi$ at  $\varphi \approx \mu$, since $\phi$ becomes heavy and it is $\chi$ that
is light. Then the potential is simply given by $m^2 |\chi|^2$, which clearly shows
that the potential is minimized at $\chi = 0$ (or $\varphi = \mu$).

Now let us discuss other cosmological issues. In order to have successful cosmology, it is necessary
to generate a right amount of baryon asymmetry and dark matter. Since the
baryonic/CDM isocurvature density perturbation is tightly constrained by observations,
it also limits possible baryogenesis and dark matter candidates~\cite{Hamaguchi:2003dc}. 
 If the baryon asymmetry is generated (or dark matter density is fixed) before the curvaton dominates the Universe,
  too large isocurvature
perturbation will be produced. Thus, both baryon asymmetry and dark matter must be generated
after the curvaton domination. 
The Hubble parameter at the curvaton domination~$H_{\rm dom}$ depends on
the reheating temperature as well as on the curvaton parameters, hence
the value of~$H_{\rm dom}$ is not uniquely determined at each point in
Figure~\ref{fig:hilltoppNG}. Largest values for~$H_{\rm dom}$ at each
point are realized when $t_{\mathrm{reh}} \leq
t_{\mathrm{osc}}$,\footnote{One can check that $H_{\mathrm{dom}}$
becomes independent of the reheating temperature when $t_{\mathrm{reh}} \leq
t_{\mathrm{osc}}$.} in such case $H_{\rm dom}$ increases as $m_\sigma$ and $f$. 
For $m_\sigma = 10\,$TeV $\hyphen 100\, \mathrm{PeV}$ and  $f \sim \GEV{17}$, 
it ranges from $10$\,GeV to 100\,TeV. 
There are several baryogenesis mechanisms which work 
at a Hubble parameter below $H_{\rm dom}$. For instance, in the Affleck-Dine 
mechanism~\cite{Affleck:1984fy,Dine:1995uk},
the baryon number is generated and fixed
when the Hubble parameter is comparable to the soft mass of the flat direction in the MSSM.
For the sfermion masses of order $10 \hyphen 100$\,TeV, it is possible  that the AD field starts
to oscillate after the curvaton domination. Since the mass of the AD field at large field value has rather large uncertainty, 
$H_{\rm dom}$ below TeV may be also allowed;
for instance, this is the case if the potential of the AD field
becomes flatter at large fields values.  There are many dark matter candidates. Since the curvaton
decays just before BBN for the case of our interest, there is an entropy dilution. One of the plausible
dark matter candidates is the QCD axion, which starts to oscillate when the plasma temperature
drops down to the QCD scale. There may be other ultralight axions which contribute
to the dark matter density. Also, the thermal relic abundance of the WIMPs as well as
WIMPs non-thermally produced by the curvaton decays are candidates for the dark matter. 

We have assumed that the curvaton is responsible for the observed density perturbation. 
From the minimalistic point of view, of course, the quantum fluctuation of the inflaton is the
leading candidate. However, requiring both an extremely flat potential for sufficiently long 
inflation and the normalization of density perturbation may be too strong constraint 
on the inflation sector. If there are many other light scalars in nature, it might be more probable
that there are two scalars, namely, the inflaton and the curvaton,  responsible for the inflationary 
expansion and the origin of density perturbations, respectively.

\section{Conclusions}
\label{sec:Conc}
In this paper we have studied non-Gaussianity of the density perturbation generated by the axionic curvaton,
focusing on the case that the curvaton initially sits near the hilltop of the potential
during inflation, and dominates the Universe before it decays. Interestingly, we have found that the non-Gaussianity parameter $f_{\rm NL}$
is positive and gets enhanced up to $30 $ (or $40$ for early reheating)
in the hilltop limit, even when the 
curvaton dominates the Universe. We have confirmed that this conclusion
holds for $n_s = 0.94 \hyphen 0.99$. 
It was also shown that in extreme cases where the axionic curvaton drives a
secondary inflation, then the produced non-Gaussianity is 
typically $f_{\mathrm{NL}} \lesssim 10$ and is smaller than
non-inflating cases. 
Note that, as long as the curvaton dominates the Universe,
$f_{\rm NL}$ cannot be larger than $30 \hyphen 40$; this should be contrasted to other
scenarios which can generate arbitrarily large non-Gaussianity, and some parameters
must be tuned to realize $f_{\rm NL} = {\cal O}(10)$. We have also pointed out that one of the plausible
candidates for the axionic curvaton is the string axion with mass of order $10 \hyphen 100$\,TeV and 
decay constant of  $\GEV{16 \hyphen 17}$. If there are many axions in the Universe, one of them
may be indeed responsible for the origin of the density perturbation.

\section*{Acknowledgements}
This work was supported by the Grant-in-Aid for Scientific Research on Innovative
Areas (No.24111702[FT], No. 21111006[FT,MK], and No.23104008[FT]) ,
Scientific Research (A) (No. 22244030 and No.21244033 [FT]), Scientific
Research  (C) (No. 14102004 [MK]) and JSPS Grant-in-Aid for Young
Scientists (B) (No. 24740135) [FT]. This work was also supported by
World Premier International Center Initiative (WPI Program), MEXT, Japan.


\appendix

\section{Inflating Curvatons}
\label{app:infcurv}

In this appendix we consider the possibility that the curvaton drives a
second inflationary stage before it starts to oscillate. 
After giving general discussions on density perturbations sourced by
such inflating curvatons, we study the case for axionic curvatons. 

\subsection{Density Perturbations from Inflating Curvatons}

\begin{figure}[htbp]
   \begin{center}
  \includegraphics[width=.46\linewidth]{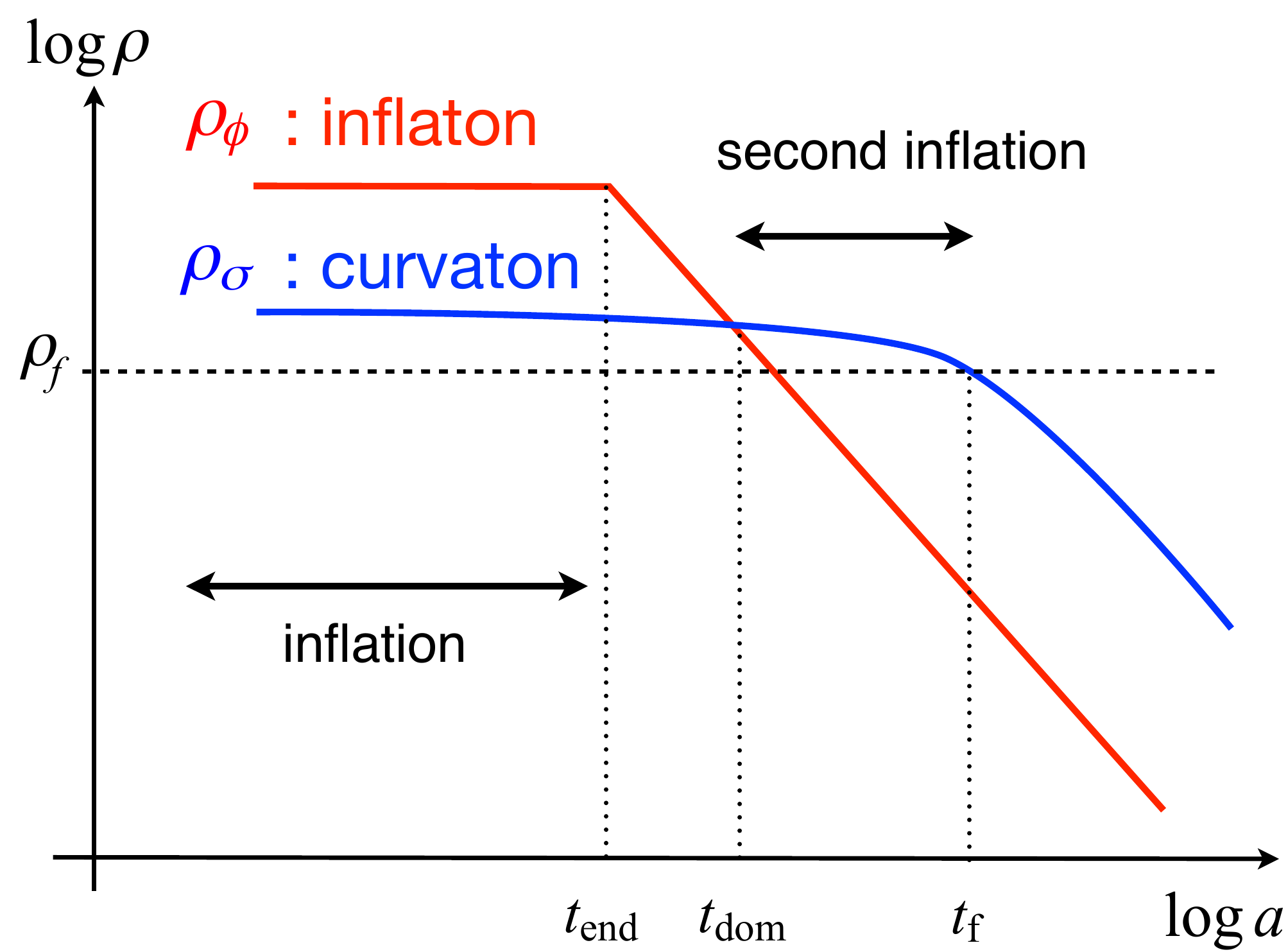}
   \end{center}
  \caption{Schematic of the time variation of energy densities in an
  inflating curvaton scenario.}
  \label{fig:schematic}
 \end{figure}

The case studied in this appendix is illustrated in
Figure~\ref{fig:schematic}: The curvaton initially (i.e. during the
first inflationary era) has negligibly tiny energy density compared to
the total energy, however dominates the universe before it starts its
oscillation. We suppose that this second inflationary period is not so 
long, and the CMB scale exits the horizon during the first inflation.
The curvaton's field fluctuations obtained from the first inflation 
lead to slight difference in the lengths of the second inflationary
periods among different patches of the universe, thus generate the
density perturbations. We also note that the second inflation is not
necessarily a slow-roll one, but may be a rapid-roll 
inflation~\cite{Linde:2001ae,Boubekeur:2005zm,Kinney:2005vj,Tzirakis:2007bf,Kobayashi:2009nv},
depending on the curvaton potential.\footnote{Note that the curvaton
field fluctuations at the CMB scale is generated during the first
inflation, thus the resulting density perturbations can be (nearly)
scale-invariant even if the second inflation is a rapid-roll one.}

Upon calculating the density perturbations using the 
$\delta \mathcal{N}$-formalism, we assume that the second
inflation lasts long enough (say, more than one e-fold) such
that during this period the inflaton energy density 
becomes negligibly tiny and then the universe is well described as
composed only of the curvaton.
This assumption allows us to choose the final uniform energy 
hypersurface to possess energy density~$\rho_f$
equal to or larger than that at the end of the second
inflationary period, cf. Figure~\ref{fig:schematic}. 
In other words, we set the final hypersurface to be before the end of
the curvaton-driven inflation, but late enough so that after which
the inflaton can be ignored and no further $\delta \mathcal{N}$ is
produced. 

The energy density of the inflaton~$\phi$ is considered to redshift as
$\rho_\phi \propto a^{-3}$ after the first inflation, and we first study
the case where the 
inflaton decay happens after the curvaton domination. 
Moreover, the curvaton~$\sigma$ is assumed to drive slow/rapid-roll
inflation after dominating the universe.
Hereafter we use the subscripts $*$ to denote values when the CMB scale
exits the horizon (during the first inflation), 
``end'' for values at the end of inflation, 
``dom'' for when the curvaton starts to dominate the universe
(i.e. $\rho_{\phi\, \mathrm{dom} } = \rho_{\sigma \, \mathrm{dom}} $), 
and ``$f$'' at the final constant energy density hypersurface. Then in
order to compute the resulting density perturbations, we would like to
obtain the $\sigma_*$-dependence of the e-folding number from the end
of inflation until the final surface ($\rho_\sigma$ is negligibly tiny
during the first inflation, thus the curvaton has little effect on the
expansion history before~$t_{\mathrm{end}}$):
\begin{equation}
 \mathcal{N} = \mathcal{N}_a + \mathcal{N}_b,
\end{equation}
where
\begin{equation}
 \mathcal{N}_a \equiv \int^{t_{\mathrm{dom}}}_{t_{\mathrm{end}}} H dt,
  \qquad
 \mathcal{N}_b \equiv \int^{t_f}_{t_{\mathrm{dom}}} H dt.
\end{equation}
Here $H = \dot{a} / a$, with an overdot denoting a time-derivative.

We take $V(\sigma)$ to be the energy
potential of the curvaton, which we assume to have no explicit time
dependence. Then using $\rho_{\sigma\, \mathrm{dom}} \simeq
V(\sigma_{\mathrm{dom}})$, one finds
\begin{equation}
  \mathcal{N}_a = \frac{1}{3} 
 \ln \frac{\rho_{\phi\, \mathrm{end}}}{\rho_{\phi \, \mathrm{dom}}}
\simeq \frac{1}{3} \ln
   \frac{\rho_{\phi\, \mathrm{end}}}{V(\sigma_{\mathrm{dom}})} ,
\end{equation}
thus
\begin{equation}
 \frac{\partial \mathcal{N}_a}{\partial \sigma_*} \simeq 
 - \frac{1}{3}
 \frac{V'(\sigma_{\mathrm{dom}})}{V(\sigma_{\mathrm{dom}})}
 \frac{\partial \sigma_{\mathrm{dom}}}{\partial \sigma_* },
\label{partialNa}
\end{equation}
where a prime denotes a derivative in terms of~$\sigma$. 
However, we will soon see that $\delta \mathcal{N}_a$ only gives a minor
contribution to the density perturbations. 

After the curvaton domination, for simplification, we 
ignore $\rho_{\phi}$ and describe the second inflation as a
single-component slow/rapid-roll inflation.\footnote{This approximation
is valid as long as the main contribution to $\delta
\mathcal{N}$ comes from the difference in the duration of the second
inflation.\label{foot:1}} 
Then the inflationary dynamics is approximated by (cf. appendix
of~\cite{Kobayashi:2009nv}),
\begin{equation}
 3 M_p^2 H^2 \simeq V,
\qquad
 \tilde{c} H \dot{\sigma}  \simeq -V',
\qquad \mathrm{where}\quad 
\tilde{c} = \frac{3 + \sqrt{9 - 12 \eta }}{2}, \quad
 \eta \equiv M_p^2 \frac{V''}{V} ,
\label{rapid-approx}
\end{equation}
which are stable attractors under the condition
\begin{equation}
 \epsilon \equiv \frac{M_p^2}{2} \left(\frac{V'}{V} \right)^2 \ll 1,
 \label{epsilon}
\end{equation}
and for a nearly constant $\eta$ satisfying $\eta \leq 3/4$. 
Here, note that $\eta$ is not bounded from below, and that $\tilde{c} \geq
3/2$. The familiar slow-roll approximations are recovered when $|\eta |
\ll 1$.
To be precise, inflation can happen even for $\epsilon > 1$ given a
large~$\tilde{c}$ (i.e. largely negative~$\eta$).\footnote{The study on 
the stability of the rapid-roll attractor given in the appendix
of~\cite{Kobayashi:2009nv} mainly considers $\tilde{c} = \mathcal{O}(1)$, but one
can easily extend their discussions to cases with $\tilde{c} \gg 1$.} 
However in this appendix we limit our studies to curvaton potentials
satisfying~(\ref{epsilon}) at $\sigma = \sigma_{\mathrm{dom}}$, as the
hilltop potentials which are 
discussed in the next section satisfy this condition. 
Then, since the curvaton field value is monotonically increasing or
decreasing in terms of time, we can use $\sigma$ as a clock,
\begin{equation}
 \mathcal{N}_b = \int^{\sigma_f}_{\sigma_{\mathrm{dom}}}
  \frac{H}{\dot{\sigma}} d \sigma .
\end{equation}
Here, note that $H/ \dot{\sigma}$ is a function of $\sigma$, and since
$\rho_f$ is a constant among different patches of the universe, so
is $\sigma_f$.\footnote{Considering $\rho_{f} \simeq V(\sigma_f)$, 
then for a potential that monotonically increases or decreases in terms
of~$\sigma$ during inflation, one sees that $\sigma_f$ is a constant.} 
Hence by partially differentiating both sides in terms of
$\sigma_*$, one obtains
\begin{equation}
 \frac{\partial \mathcal{N}_b}{\partial \sigma_*} \simeq 
 \left. \frac{\tilde{c} V}{3 M_p^2 V'} \right|_{\sigma = \sigma_{\mathrm{dom}}}
 \frac{\partial \sigma_{\mathrm{dom}}}{\partial \sigma_*} .
\label{partialNb}
\end{equation}

In order to compute $\partial \sigma_{\mathrm{dom}} / \partial \sigma_*$, 
we make use of the slow-roll approximation $3 H \dot{\sigma} \simeq -V'$
while $t \leq t_{\mathrm{end}}$. 
During $ t_{\mathrm{end}} \leq t \leq t_{\mathrm{dom}}$, 
for simplification we treat the universe
as a matter dominated one and adopt $\frac{9}{2} H \dot{\sigma} \simeq
-V'$ (cf. (\ref{CDapp}), see also Footnote~\ref{foot:1}).
Also using $\dot{H} / H^2 = -3/2$ for $ t_{\mathrm{end}} \leq t \leq
t_{\mathrm{dom}}$, then one can check that
\begin{equation}
 \int^{\sigma_{\mathrm{dom}}}_{\sigma_*} \frac{d \sigma}{V'(\sigma)}
 \simeq \frac{4}{27} \int^{H_{\mathrm{dom}}}_{H_{\mathrm{end}}}
 \frac{dH}{H^3} + (\mathrm{terms\, \, independent \, \, of\, \,
 }\sigma_*).
\label{A.9}
\end{equation}
Partially differentiating both sides by $\sigma_*$, and
using $3 M_p^2 H_{\mathrm{dom}}^2 \simeq 2 V
(\sigma_{\mathrm{dom}})$,\footnote{This may seem contradicting with the 
slow/rapid-roll approximation~(\ref{rapid-approx}) $3 M_p^2 H^2 \simeq
V$, but the numerical coefficient of~$V$ only affects a $M_p^2 (V'/V)^2$
term which is dropped in the final expression~(\ref{eq10}), thus we will
not worry about it.} one obtains
\begin{equation}
 \frac{\partial \sigma_{\mathrm{dom}}}{\partial \sigma_*} \simeq 
 \frac{V'(\sigma_{\mathrm{dom}})}{V'(\sigma_*)} ,
\label{eq10}
\end{equation}
where we have dropped the contribution from the right hand side of
(\ref{A.9}) from the condition (\ref{epsilon}) satisfied at $\sigma =
\sigma_{\mathrm{dom}}$. 

\vspace{\baselineskip}

Combining the above results, we can calculate the density perturbation spectrum:
\begin{equation}
 \mathcal{P}_\zeta = \left( \frac{\partial \mathcal{N}}{\partial
      \sigma_*} \right) \left( \frac{H_*}{2 \pi  }\right)^2,
\label{A.11}
\end{equation}
where (again using (\ref{epsilon}))
\begin{equation}
 \frac{\partial \mathcal{N}}{\partial \sigma_*} 
\simeq
 \frac{\tilde{c}(\sigma_{\mathrm{dom}}) V(\sigma_{\mathrm{dom}})}{3 M_p^2
 V'(\sigma_*)} .
\label{A.12}
\end{equation}
The spectral index follows as 
\begin{equation}
 n_s -1 \simeq  2 \frac{\dot{H}_*}{H_*^2} + \frac{2}{3}
  \frac{V''(\sigma_*)}{H_*^2},
\label{A.13}
\end{equation}
taking the same form as for non-inflating curvatons (\ref{eq:ns}). 
The non-Gaussianity parameter can also be calculated:
\begin{equation}
 f_{\mathrm{NL}} = \frac{5}{6} \frac{\partial^2 \mathcal{N}}{\partial
  \sigma_*^2}  \left( \frac{\partial \mathcal{N}}{\partial \sigma_*}
	       \right)^{-2}  
\simeq \frac{5  }{2 \tilde{c} (\sigma_{\mathrm{dom}})}
 \left\{
  \left(
\frac{M_p V'(\sigma_{\mathrm{dom}})}{V(\sigma_{\mathrm{dom}})} 
\right)^2 
 -   \frac{M_p^2 V''(\sigma_*)}{V(\sigma_{\mathrm{dom}}) }
\right\}
 + \cdots ,
\label{eq:fNL}
\end{equation}
where $\cdots$ denotes terms proportional to $\partial \tilde{c}
(\sigma_{\mathrm{dom}}) / \partial \sigma_{\mathrm{dom}}$. 
One immediately sees that the first term in the $\{\, \}$ parentheses 
is much smaller than unity from the condition~(\ref{epsilon}), while the
second term can be larger than unity for rapid-roll inflation, i.e.
$|\eta | \gtrsim 1$.

\vspace{\baselineskip}

In the above discussion, we have considered the inflaton to decay after
the curvaton domination. 
Similar computations can be carried out also for the case where the
inflaton decays between the first and second inflationary periods,
by approximating the universe as matter dominated while
$t_{\mathrm{end}} \leq t \leq t_{\mathrm{reh}}$
(here the subscript ``reh'' denotes values at $H=
\Gamma_\phi$, when the inflaton is assumed to suddenly decay),
and then radiation dominated while $t_{\mathrm{reh}} \leq t \leq
t_{\mathrm{dom}}$. 
Further assuming the condition (\ref{epsilon}), and also that 
the tilt of the curvaton potential at $\sigma_{\mathrm{reh}}$ to be not
much greater than at~$\sigma_{\mathrm{dom}}$, i.e.,
\begin{equation}
 \left| V'(\sigma_{\mathrm{reh}}) \right| \lesssim 
\left| V'(\sigma_{\mathrm{dom}}) \right|,
\label{condreh}
\end{equation}
then one obtains the same results (\ref{A.12}), (\ref{A.13}), and
(\ref{eq:fNL}). 
Whether the inflaton decays before or after the curvaton domination has
little effect since the density perturbations are sourced mainly through
different patches of the universe experiencing slightly longer/shorter
periods of the second inflation.

\vspace{\baselineskip}

In summary, independently of whether the inflaton decays before/after
the curvaton domination, 
under the condition (\ref{epsilon}) (and also (\ref{condreh}) for
$t_{\mathrm{reh}} < t_{\mathrm{dom}}$),
the linear perturbation sourced by inflating curvatons is of the form
(\ref{A.11}) with (\ref{A.12}), the spectral index is (\ref{A.13}), and
the non-linearity parameter is given by (\ref{eq:fNL}).

\subsection{Inflating at the Hilltop}
\label{app:inf_hilltop}

As an example, let us consider inflating curvatons with a hilltop
potential 
\begin{equation}
 V(\sigma) = V_0 - \frac{1}{2}m^2 (\sigma - \sigma_0)^2 ,
\label{hilltoppot}
\end{equation}
where $V_0$, $m$, and $\sigma_0$ are constants. Given that the curvaton
is located sufficiently close to the hilltop~$\sigma_0$ such that 
$V_0 \gg m^2 (\sigma - \sigma_0)^2$ and $V_0^2 \gg M_p^2 m^4 (\sigma -
\sigma_0)^2$ until the curvaton starts driving inflation,
then one finds 
\begin{equation}
 \frac{\partial \mathcal{N}}{\partial \sigma_*} \simeq
 -  \frac{3 + \sqrt{9 - 12 \eta} }{6 \eta } \frac{1}{ (\sigma_0 - \sigma_*)},
\label{A.17}
\end{equation}
and 
\begin{equation}
 f_{\mathrm{NL}} \simeq -  \frac{5 \eta}{2 \tilde{c}}
 = \frac{5}{12} \left( -3 + \sqrt{9 - 12 \eta}  \right),
\label{A.18}
\end{equation}
where
\begin{equation}
\eta \simeq -  \frac{M_p^2 m^2}{V_0}.
\end{equation}
Note especially that $\eta$ is (almost) a constant which is negative in this example. 
We also remark that  we have dropped the contribution on~$f_\mathrm{NL}$
from the first term in the parentheses of (\ref{eq:fNL}) which is
clearly smaller than unity.

\subsection{Axionic Curvatons}

In this final subsection, we look into axionic curvatons inflating at
the hilltop of the potential~(\ref{eq:sin}). 
(We do not consider axionic curvatons away from the hilltop driving
large-field inflation with super-Planckian decay constants~$f$.)
The discussions in Section~\ref{app:inf_hilltop} can be applied to this
case by simply substituting 
\begin{equation}
 V_0 = 2 \Lambda^4, \quad
 m^2 = \frac{\Lambda^4}{f^2} , \quad
 \sigma_0 = f \pi . 
\end{equation}
Then one can see that the negative $ \eta$ parameter is determined
merely by the symmetry breaking scale~$f$ as
\begin{equation}
 \eta \simeq - \frac{M_p^2}{ 2 f^2} .
\label{etaforpNG}
\end{equation}

\begin{figure}[htbp]
 \begin{minipage}{.48\linewidth}
  \begin{center}
 \includegraphics[width=\linewidth]{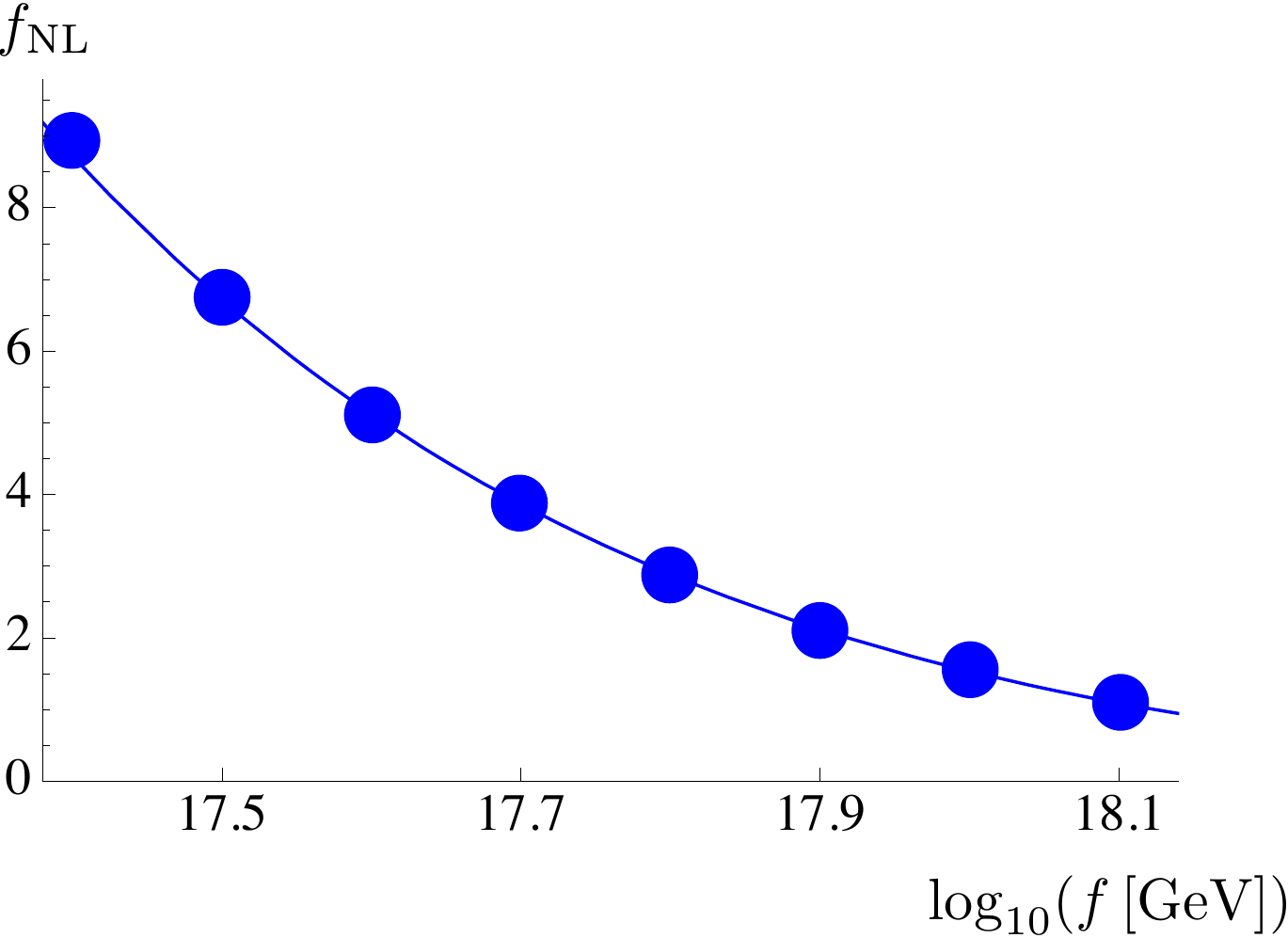}
  \end{center}
 \end{minipage} 
 \begin{minipage}{0.01\linewidth} 
  \begin{center}
  \end{center}
 \end{minipage} 
 \begin{minipage}{.48\linewidth}
  \begin{center}
 \includegraphics[width=\linewidth]{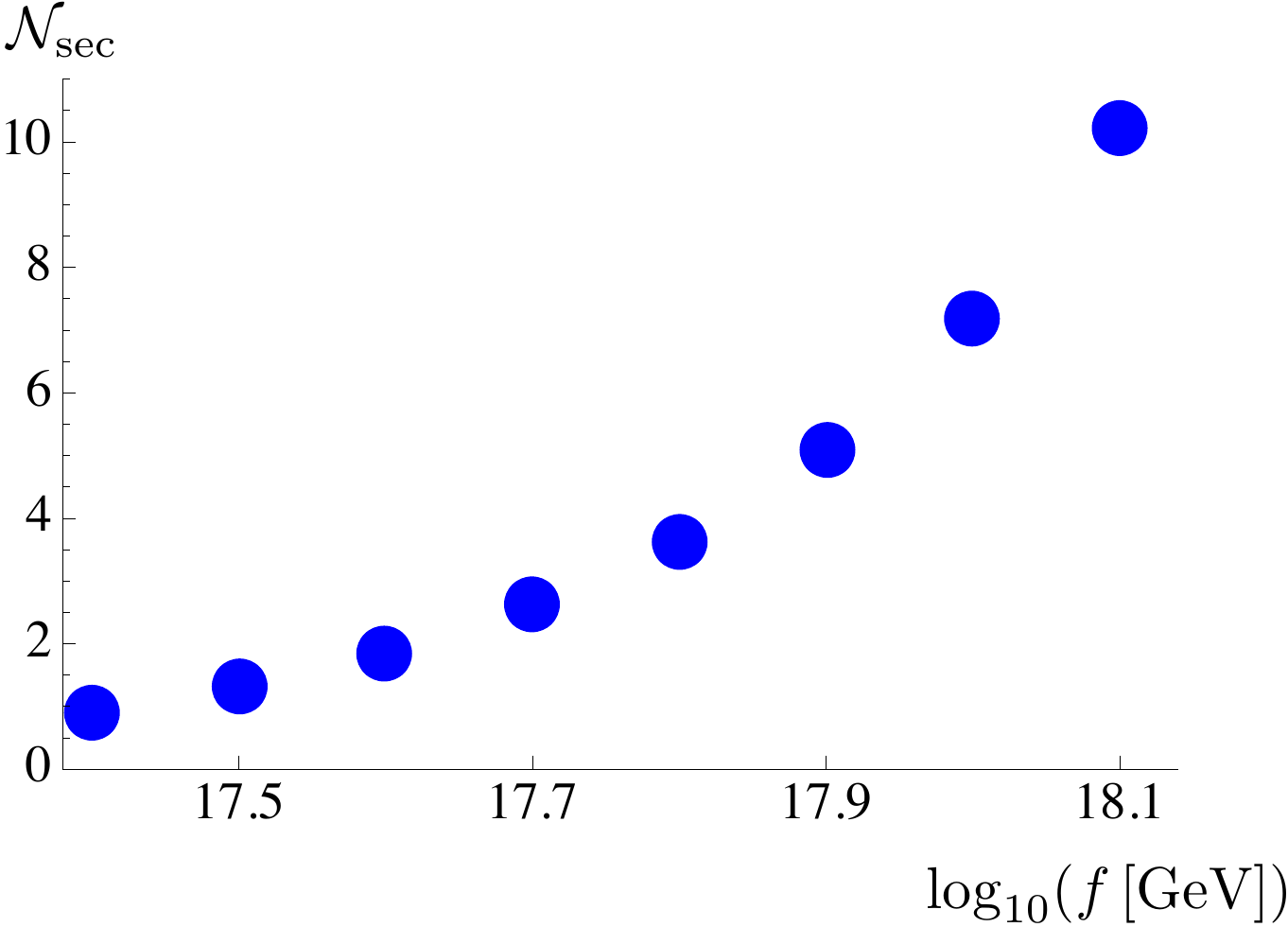}
  \end{center}
 \end{minipage} 
  \caption{Left: Non-Gaussianity from an inflating axionic curvaton with a fixed
 mass $ m_\sigma =  10^8\, \mathrm{GeV}$, when increasing the decay
 constant~$f$ beyond the right blue edge of the allowed parameter window
 in Figure~\ref{fig:hilltoppNG}. Parameters other than $m_\sigma$ and
 $f$ are fixed from requiring $\mathcal{P}_\zeta \approx 2.4 \times
 10^{-9}$ and $n_s \approx 0.96$. Right: Number of e-folds in the second
 inflationary period driven by the axionic curvaton.}
  \label{fig:inflating_pNG}
\end{figure}

Non-Gaussianities from inflating axionic curvatons are in general 
smaller compared to non-inflating cases, which can be seen
from~(\ref{A.18}): Large~$f_{\mathrm{NL}}$ 
requires a large, negative~$\eta$, however a large~$|\eta |$
substantially accelerates the curvaton thus makes it challenging for the
curvaton to drive inflation in the first place. 
We show this explicitly in Figure~\ref{fig:inflating_pNG}, which
investigates the parameter space of axionic curvatons beyond the allowed
window for non-inflating axionic curvatons discussed in
Section~\ref{sec:pNGcurvaton}. 
Here we fix the curvaton mass to $ m_\sigma =  \Lambda^2 / f = 10^8\,
\mathrm{GeV}$, and investigate large~$f$ values beyond the 
right blue edge (corresponding to the requirement that
the curvaton be subdominant until it starts oscillating)
in Figure~\ref{fig:hilltoppNG}. 
Using (\ref{A.13}), we fix the (first) inflation
scale~$H_{\mathrm{inf}}$ from the spectral index $n_s \approx 0.96$
assuming constant~$H$ during inflation, and also fix the curvaton
position at CMB scale horizon exit $\sigma_* / f $ from
$\mathcal{P}_\zeta \approx 2.4 \times 10^{-9}$ using (\ref{A.17}).  
Furthermore, we set the number of e-foldings between the CMB scale
horizon exit and the end of the (first) inflation as $\mathcal{N}_* =
50$, the curvaton decay rate~$\Gamma_\sigma$ by (\ref{decayrate}) with
$\beta = 1$, and the inflaton decay rate~$\Gamma_\phi$ small enough such
that the inflaton decays after the curvaton domination. (The explicit
value of~$\Gamma_\phi$ is irrelevant for the density perturbations,
however whether the inflaton decays before/after the curvaton domination
slightly affects the number of e-folds obtained in the second
inflation.) The resulting  
non-Gaussianity~$f_{\mathrm{NL}}$ is plotted as a function of~$f$ in
Figure~\ref{fig:inflating_pNG}, where the blue solid line denotes the
analytic calculation (\ref{A.18}) with (\ref{etaforpNG}). We have
also numerically computed~$f_{\mathrm{NL}}$, whose results are shown as
blue dots in the figures. One sees that the analytic and numerical
results match well. In the right figure, we also show the number of
e-folds~$\mathcal{N}_{\mathrm{sec}}$ obtained in the second inflationary
period driven by the axionic curvaton. $\mathcal{N}_{\mathrm{sec}}$ here is
defined as the e-folding number from the curvaton domination until when
the curvaton starts oscillating, i.e. (\ref{conditionX}).
When the second inflationary period is very short, the analytic
estimations derived in this appendix are invalid, which sources the
slight difference between the analytic and numerical computations 
of~$f_{\mathrm{NL}}$ at $f \approx 10^{17.4}\, \mathrm{GeV}$. 
When further increasing~$f$ beyond the plotted regime, $f_{\mathrm{NL}}$
becomes further suppressed while $\mathcal{N}_\mathrm{sec}$ rapidly
increases, soon making the axionic curvaton responsible for driving
most of the inflationary e-folds after the CMB scale horizon exit. 
In summary, non-Gaussianity in the region beyond the right
blue edge in Figure~\ref{fig:inflating_pNG} decreases for
larger~$f$, taking values smaller than $\sim 10$ in most of the region. 
Larger non-Gaussianity is generated when closer to the edge, i.e. when
the second inflationary period is very short and the situation is close
to the familiar non-inflating curvatons.

Let us also note that the power spectrum (\ref{A.17}) is now written as 
\begin{equation}
 \mathcal{P}_{\zeta}^{1/2} \simeq \kappa 
 \left(1 - \frac{\sigma_*}{\pi f}   \right)^{-1}  \frac{H_*}{M_p},
 \qquad \mathrm{where} \quad 
 \kappa \equiv \frac{3 + \sqrt{9 - 12 \eta} }{6 \pi^2
  \sqrt{-2 \eta }  }  . 
\end{equation}
For a sub-Planckian~$f$ (i.e. $f \leq M_p$), the prefactor $\kappa$ can
only take values within $0.04 \lesssim \kappa \lesssim
0.12$. Therefore, once the initial position of the curvaton~$\sigma_* /
\pi f$ is given, the inflationary scale~$H_*$ needs to be tuned to a
rather narrow scale range in order for an inflating axionic curvaton to
source the linear perturbation with an appropriate amplitude. This is in
contrast to non-inflating axionic curvatons, which can work with a wide
range of inflationary scales for each value of $\sigma_* / \pi 
f$~\cite{Kawasaki:2011pd}. 

We remark that the inflating curvaton was studied in Ref.~\cite{Dimopoulos:2011gb}, however
their results differ from ours.  In particular, they claimed that the non-Gaussianity parameter
$f_{\rm NL}$ is {\it negative} and of order unity~\footnote{They assumed
the curvaton mass to be much lighter than the Hubble parameter during
inflation, but even in this case the analysis shown in
Figure~\ref{fig:inflating_pNG} does not change much since $f_{\rm NL}$ from
an axionic curvaton is set merely by the decay constant~$f$.}, while we
have shown that $f_{\rm NL}$ from a hilltop curvaton is {\it positive} 
and can take values larger as well as smaller than order unity. 
Moreover  we have confirmed our results for the case of axionic curvatons
by numerical calculations as shown in Fig.~\ref{fig:inflating_pNG},
where one sees that $f_{\mathrm{NL}}$ varies from $9$ to $1$ as the
e-folding number in the second inflation increases from $1$ to $10$.

\end{document}